\documentclass[useAMS,fleqn,usenatbib]{mnras}
\usepackage[T1]{fontenc}
\usepackage{ae,aecompl}
\usepackage{bethmacros}
\usepackage{graphicx}
\usepackage{amssymb}
\usepackage{amsmath}
\usepackage{times}

\def\spose#1{\hbox to 0pt{#1\hss}}
\def\lta{\mathrel{\spose{\lower 3pt\hbox{$\mathchar"218$}}
     \raise 2.0pt\hbox{$\mathchar"13C$}}}
\def\gta{\mathrel{\spose{\lower 3pt\hbox{$\mathchar"218$}}
     \raise 2.0pt\hbox{$\mathchar"13E$}}}

\title[Globular Cluster Formation Environments]{Where did the globular clusters
of the Milky Way form? Insights from the E-MOSAICS simulations}

\author[Keller et al.]{Benjamin W. Keller$^1$\thanks{Email: benjamin.keller `at'
uni-heidelberg.de},  J. M. Diederik Kruijssen$^1$, Joel Pfeffer$^2$, 
Marta Reina-Campos$^1$, \newauthor Nate Bastian$^2$, Sebastian Trujillo-Gomez$^1$, Meghan
E.  Hughes$^2$, Robert A. Crain$^2$ 
\vspace*{6pt}\\
$^1$Astronomisches Rechen-Institut, Zentrum f{\"u}r Astronomie der Universit\"at
Heidelberg, M{\"o}nchhofstra{\ss}e 12-14, D-69120 Heidelberg, Germany \\ 
$^2$Astrophysics Research Institute, Liverpool John Moores University, 146
Brownlow Hill, Liverpool L3 5RF, UK}

\begin{document}
\maketitle
\label{firstpage}
\begin{abstract} 
    Globular clusters (GCs) are typically old, with most having formed at
    $z\gtrsim2$.  This makes understanding their birth environments difficult,
    as they are typically too distant to observe with sufficient angular
    resolution to resolve GC birth sites.  Using 25 cosmological zoom-in
    simulations of Milky Way-like galaxies from the E-MOSAICS project, with
    physically-motivated models for star formation, feedback, and the formation,
    evolution, and disruption of GCs, we identify the birth environments of
    present-day GCs.  We find roughly half of GCs in these galaxies formed
    in-situ ($52.0 \pm 1.0$ per cent) between $z\approx2-4$, in turbulent,
    high-pressure discs fed by gas that was accreted without ever being strongly
    heated through a virial shock or feedback.  A minority of GCs form during
    mergers ($12.6\pm 0.6$ per cent in major mergers, and $7.2\pm 0.5$ per cent
    in minor mergers), but we find that mergers are important for preserving the
    GCs seen today by ejecting them from their natal, high density interstellar
    medium (ISM), where proto-GCs are rapidly destroyed due to tidal shocks from
    ISM substructure.  This chaotic history of hierarchical galaxy assembly acts
    to mix the spatial and kinematic distribution of GCs formed through
    different channels, making it difficult to use observable GC properties to
    distinguish GCs formed in mergers from ones formed by smooth accretion, and
    similarly GCs formed in-situ from those formed ex-situ.  These results
    suggest a simple picture of GC formation, in which GCs are a natural outcome
    of normal star formation in the typical, gas-rich galaxies that are the
    progenitors of present-day galaxies.
\end{abstract}

\begin{keywords}
-- galaxies: formation --  galaxies: evolution -- galaxies: haloes -- galaxies: star
formation -- galaxies: star clusters: general -- globular clusters: general
\end{keywords}

\section{Introduction}
A ubiquitous feature of galaxies in the nearby universe are their populations of
globular clusters (GCs).  These old ($\tau\sim 10\Gyr$), massive
($M\sim10^4-10^6\Msun$) stellar clusters
\citep[e.g.][]{Brodie2006,Kruijssen2014b,Forbes2018} are found distributed
throughout the haloes of nearly all galaxies with $M_* \gtrsim 10^9\Msun$
\citep{Harris2017b}.  The ages of these objects tell us that many of them formed
near cosmic noon, at $z\sim2$
\citep[e.g.][]{Forbes2010,Dotter2010,Dotter2011,VandenBerg2013,Kruijssen2019a,Reina-Campos2019},
when the cosmic star formation rate was at its peak \citep{Madau2014}.  Age
determinations have yet to reach the level of precision, however, where
observations alone can tell us the precise time line of GC formation in the
Milky Way (MW) or elsewhere.  The GC populations we see in galaxies across many
decades of halo and stellar mass show remarkable differences compared to the
``normal'' field stellar populations of those same galaxies.  GCs are typically
older \citep{MarinFranch2009,VandenBerg2013}, more metal poor
\citep{Puzia2005,Sarajedini2007}, and broadly distributed through the halo
compared to field stars \citep{Zinn1985}.  Both the number \citep{Blakeslee1997}
and total mass \citep{Spitler2009,Harris2017b} of GCs appears to be a constant
ratio of halo mass, unlike the total stellar mass, which both abundance matching
\citep{Behroozi2013,Moster2013} and weak lensing studies \citep{Hudson2015} have
confirmed to be a non-linear function of halo mass.  In different galaxies,
unimodal \citep[e..g][]{Harris2017a}, bimodal \citep[e.g.][]{Peng2006}, and even
trimodal distributions \citep[e.g.][]{Blom2012,Usher2012} of GC metallicities
have been observed.  Understanding when, where, and how these objects form can
help us understand star formation in some of the most extreme cosmic
environments.  As the epoch of GC formation may coincide with cosmic noon
\citep{Reina-Campos2019}, understanding GC formation will help us better
understand star formation during a key phase of the Universe's evolution.

The differences between the populations of stars we see in GCs and the rest of
the field stars have prompted a critical question:  do they signal that special,
early universe physics (distinct from ``normal'' star formation) is required to
form GCs?  Over time, a number of different formation scenarios have been
proposed.  The earliest proposed mechanisms for forming GCs are wildly different
to the mechanisms we currently believe produce the vast majority of field stars.
\citet{Peebles1968} argued that as the Jeans mass of the typical
post-recombination intergalactic medium was comparable to the mass of observed
GCs ($10^5-10^6\Msun$), and thus the collapse of pre-galactic gas clouds
produced the GCs we see today.  This model cannot produce metal-rich clusters,
and would produce a cluster population with a radial distribution identical to
the dark matter halo (rather than the more centrally concentrated distribution
we observe). Naturally, the GCs produced via this mechanism would contain the
oldest stars in existence (since the formation of the remaining non-GC stellar
populations would necessarily follow the formation of galaxies).
\citet{Fall1985} proposed a mechanism that begins producing GCs once the
formation of galaxies has begun, through a two-phase instability in the hot
galactic corona.  The discovery of GCs in low mass field dwarf galaxies, which
lack a hot corona, means that at least some GCs cannot be formed through this
mechanism \citep{Larsen2012}.  Similar difficulties are faced by models which
form GCs only during major mergers \citep{Ashman1992},   as this merger-driven
scenario fails to reproduce the metallicity distributions observed both in
ellipticals \citep{Forbes1997} and the MW \citep{Griffen2010}.  Mergers
may not only trigger the formation of GCs, but instead transport GCs into the
galaxy's halo.  Finally, the nuclei of dwarf galaxies have similar masses and sizes to
GCs, and so have been proposed as the progenitors of MW GCs after the rest of
the dwarf galaxy is stripped away through tidal interactions
\citep[e.g.][]{Zinnecker1988,Torsten2008}.  Studies of the assembly history of
galaxies have suggested that there are simply not enough stripped dwarf nuclei
to account for the GC population we see today \citep{Pfeffer2014}.

The high-redshift environment that globular clusters form in has made studying
their birth conditions difficult.  The galactic ecosystem at $z>2$ was rather
different than it is today:  mergers were far more frequent
\citep{Lacey1993,Genel2009}, low virial temperatures could allow unshocked gas
to feed discs directly through cold flows along filaments
\citep{Dekel2006,Woods2014}, and discs were clumpy and irregular
\citep{Ceverino2010,Genzel2011}.  All this leads to a potential formation
environment for globular clusters that is different to that of a typical
star-forming galaxy in the local universe: violently turbulent, gas-rich,
high-pressure discs.  Despite these differences, there do exist some analogs to
these environments in the local universe (especially in merging and starbursting
galaxies) and in these environments, potential ``new'' globular clusters are
found in the form of young massive clusters (YMCs) \citep{PortegiesZwart2010}.
While YMCs have comparable mass and size to observed GCs, they typically form in
galaxies with much higher metallicity \citep{Ma2016} than the median metallicity
of GCs, and they have also not been subject to Gyrs of evolution.  Their
locations within their host galaxies are also quite different to that of GCs.
While YMCs are seen forming in the dense ISM of the present-day galaxy disc, a
significant fraction of the GC population orbits the galaxy at large
radii, in a spherical distribution through the stellar halo.  In order to
build a population of GCs in the stellar halo, these GCs must either be flung
out of the disc by merger events (an {\it in-situ} mechanism), or tidally
stripped from accreted galaxies (an {\it ex-situ} mechanism), or simply have
been formed at large radii (as in \citealt{Peebles1968}).

Regardless of the physics involved in the formation of GCs, a second selection
step obscures their birth environment: the evolution of the cluster and
dynamical disruption as it orbits within the galaxy for $\sim10\Gyr$.  Over this
time, it will experience heating and disruption through the tidal field of the
galaxy \citep{Ambartsumian1938,Spitzer1940,Henon1961,Lee1987,Baumgardt2003}, and
tidal shocks when eccentric orbits bring them through the galactic disc and
bulge \citep{Aguilar1988}.  Early in the cluster's lifespan, it may pass through
spiral arms and giant molecular clouds (GMCs), subjecting it to further intense
tidal shocks
\citep{Lamers2005,Gieles2006,Elmegreen2010a,Elmegreen2010b,Kruijssen2011,Kruijssen2015}.
Clusters within the halo may inspiral over time, as they lose angular momentum
through dynamical friction \citep{Tremaine1976}.  All of this means that the
population of GCs we see today may have little resemblance to the population of
proto-GCs that formed within the galaxy.  If these disruption processes act much
more strongly on clusters formed through one scenario we may find that despite
that scenario producing most proto-GCs, those clusters that survive to $z=0$ are
mostly formed through other mechanisms.  Realistic modelling of both the
formation and disruption physics is critical to being able to use simulations to
study the formation of GCs.

The past half century has flipped the problem of GC formation on its head.
Rather than lacking any explanation for the origin of GCs and their differences
from the population of field stars, we now have a wealth of different
theoretical models, some of which are quite successful in reproducing the
metallicity, age, and number distributions that are seen observationally.
Indeed, the problem in understanding the origin of GCs is now a question of
which mechanisms, in what environments, and with what frequency, lead to the GC
populations we see today.  In this paper, we use the E-MOSAICS suite of 25
cosmological zoom-in simulations of $L^*$ galaxies to study the birth
environment of the GC systems observed in present-day galaxies.  The E-MOSAICS
simulations are a state-of-the-art set of simulations of galaxy formation and
evolution that simultaneously model the formation of stellar clusters and their
parent galaxy with sub-grid models for star and cluster formation, stellar
feedback, and tidal disruption, along with hydrodynamics, radiative cooling, and
gravity in a fully cosmological environment.  This lets us trace back the GCs we
see at $z=0$ to determine the conditions of the gas from which they are born,
and compare this population to the progenitor clusters (which have not yet
experienced disruption effects). The purely local, yet environmentally-dependent
physical models for GC formation and evolution allow us to study the birth
environments of GCs without making assumptions about the relative importance of
mergers, the age of the universe, or global galaxy properties.

The structure of this paper is as follows.  In section~\ref{s:sims}, we detail
the numerical methods used in the E-MOSAICS simulations.
Section~\ref{s:birthenv} outlines the results we find for the birth environments
of globular clusters, and the scenarios that produce them.
Section~\ref{s:collisions} looks specifically at globular clusters potentially
produced in the collisions of substructure in the haloes of galaxies.
Section~\ref{s:discussion} places our results in the context of other
theoretical and observational studies of GC birth.  We summarise our conclusion
in Section~\ref{s:conclusions}.  

\section{The E-MOSAICS simulations}
\label{s:sims}
\begin{figure*}
    \includegraphics[width=\textwidth,keepaspectratio]{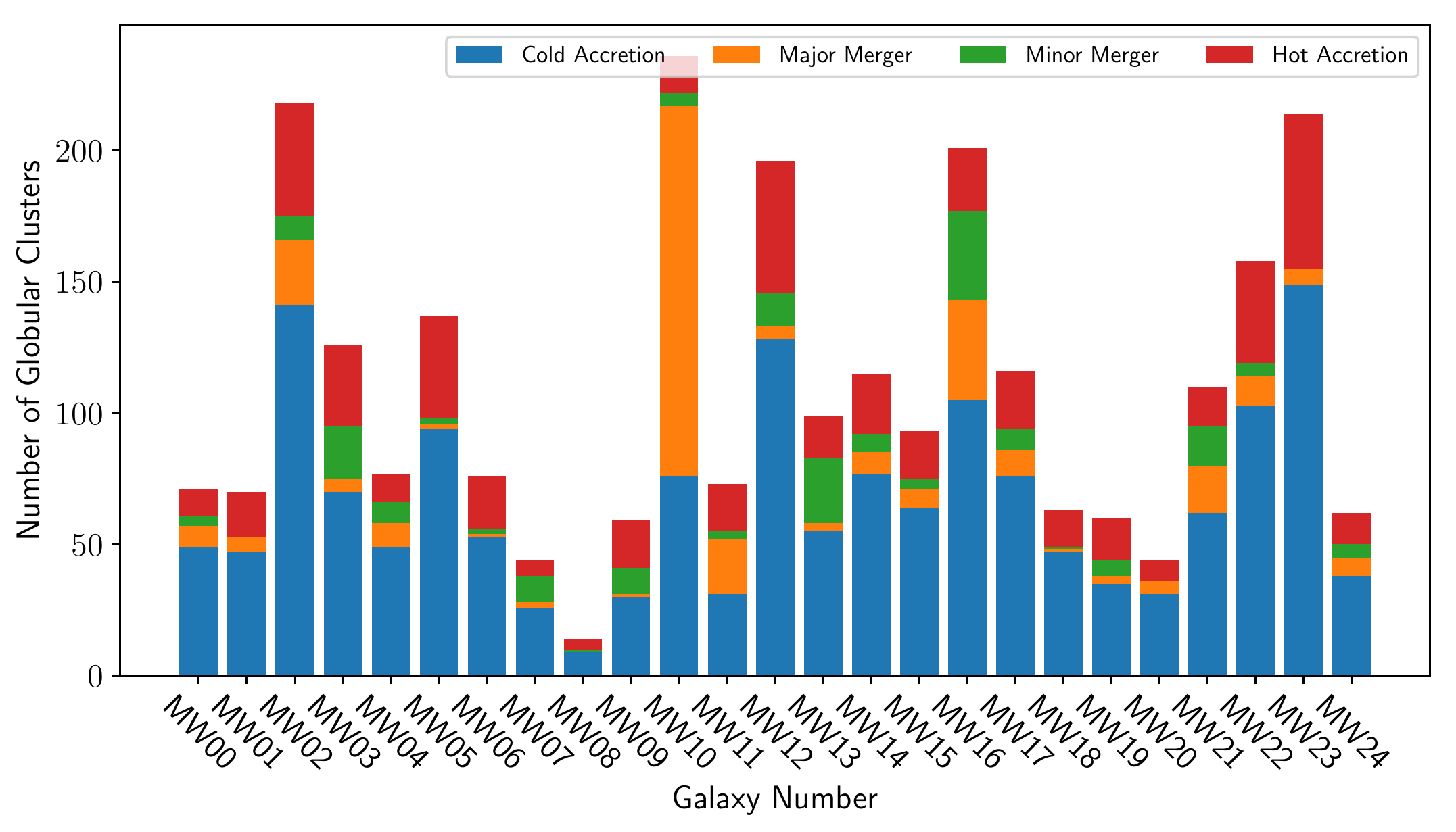}
    \caption{The formation environments of the $z=0$ GC populations in each of
    the 25 E-MOSAICS galaxies.  The majority of GCs in every E-MOSAICS galaxy,
    except for MW10 and MW11, forms from gas which is smoothly accreted while
    cold, never exceeding $2\times10^5\K$ prior to forming a GC.  Both major and
    minor mergers only contribute a small fraction of the total number of GCs
    seen.}
    \label{formation_environment}
\end{figure*}
The E-MOSAICS simulations were introduced by \citet{Pfeffer2018,Kruijssen2019a},
and described in detail in section 2 there.  In brief, E-MOSAICS consists of 25
cosmological zoom-in simulations \citep{Katz1993} of typical $L^*$, MW-like
spiral galaxies.  The simulations build on the successful EAGLE cosmological
volume simulations \citep{Schaye2015,Crain2015} by using the same set of physics
for gas cooling and heating, formation of both stars and black holes, as well as
feedback from the same.  The details of the physics model used in both EAGLE and
E-MOSAICS can be found in \citet{Crain2015} and \citet{Pfeffer2018}, the
enhanced hydrodynamics method {\sc anarchy} is detailed in \citet{Schaye2015},
and the results of the improvements are described in \citet{Schaller2015}.  The
EAGLE simulations reproduce the galactic stellar mass function and size-mass
relation \citep{Baldry2012} through careful calibration of the star formation
and feedback parameters used.  This means that other galaxy properties and
scaling relations, such as the cosmic star formation history, the black hole to
stellar mass relation, mass-metallicity relation, and others are all predictions
of the EAGLE physics model, rather than quantities that are explicitly tuned to
match observations.  The interplay of galaxy assembly and the star formation and
feedback processes then emergently reproduce the star forming sequence
\citep{Crain2015}, cosmic star formation history \citep{Furlong2015}, observed
mass-discrepancy acceleration relation \citep{Ludlow2017}, HI-stellar mass
relation \citep{Crain2017}, QSO absorption features
\citep{Oppenheimer2016,Turner2017}, and mass-metallicity relation
\citep{DeRossi2017}.

The EAGLE model uses the \citet{Wiersma2009} model for radiative cooling and
heating, assuming ionisation equlibrium and a \citet{Haardt2012} UV background.
Star formation is handled using a a pressure-dependent reformulation of the
\citet{Kennicutt1998} star formation law \citep{Schaye2004}.  The slope of the star formation
relation $\dot m_* \propto P^{(n-1)/2}$ follows the standard Kennicutt slope of
$n=1.4$ below $n_H < 10^3\hcc$, but increases to $n=2$ at higher densities.
Star formation is allowed to occur in gas which exceeds $n_H = 0.1\hcc
(Z/0.002)^{-0.64}$, where $Z$ is the local gas metallicity.
A thorough discussion of this model, a justification of its parameters, and a
comparison to simpler star formation models can be found in \citet{Schaye2015}
and \citet{Crain2015}.

The haloes simulated in E-MOSAICS are selected from the EAGLE Recal-L025N0752
volume, and re-simulated with a factor of 8 better mass resolution than the
$100\Mpc$ EAGLE volume (corresponding to a factor of 2 better spatial
resolution).  The baryonic particle mass is $2.25\times10^5\Msun$, with a
Plummer-equivalent gravitational softening length of $1.33$ comoving kpc prior
to $z=2.8$ and $350$ physical pc from that point forward.  Merger trees are
generated, as in \citet{Schaye2015} and \citet{Qu2017}, using {\sc subfind}
\citep{Springel2001} on the 29 snapshots saved between redshift 20 and 0.  

The distinguishing feature that sets E-MOSAICS apart from EAGLE, or from the
APOSTLE \citep{Sawala2016} and Cluster-EAGLE \citep{Barnes2017,Bahe2017}
zoom-in simulations, which also use the EAGLE physics model, is the inclusion
of a set of sub-grid physics models for the formation, evolution, and
disruption of gravitationally bound stellar clusters (combined in the MOSAICS
sub-grid cluster model \citealt{Kruijssen2011,Pfeffer2018}).  These
stellar cluster models are
fully local, depending only on physical properties of the local environment,
rather than disc or halo averaged properties, or non-local measurements (such
as the identification of mergers).  An in-depth discussion of these models for
GC formation and evolution is
presented in \citet{Pfeffer2018} and \citet{Kruijssen2019a}, which we briefly
summarise here.  The cluster formation model in E-MOSAICS is determined by the 
theoretical model for the cluster formation efficiency (CFE) derived in
\citet{Kruijssen2012c}, which relates the CFE to the local gas volume density,
velocity dispersion, and thermal sound speed, and reproduces the CFEs observed
in nearby star-forming galaxies.  Each star particle contains a
population of clusters with an initial cluster mass function (ICMF) described
by a \citet{Schechter1976} function, with a ICMF truncation mass related to the
maximum GMC mass \citep{Reina-Campos2017}.  This is set by the fraction of a
locally-calculated Toomre mass which can collapse and form stars prior to being
disrupted by feedback.  Cluster evolution then consists of three major
components: stellar evolution, using the same stellar evolution tracks as
EAGLE, two-body relaxation, and tidal shocks.  Two-body relaxation and tidal
disruption rates are calculated using the locally-calculated tidal tensor
\citep{Gnedin2003,Prieto2008,Kruijssen2011}, which allows on-the-fly
calculation of cluster disruption based on the tidal environment they find
themselves in.  Finally, the dynamical friction timescale for a cluster to fall
into the centre of the galaxy is calculated with a postprocessing algorithm
\citep{Pfeffer2018}.  If the dynamical friction timescale calculated is less
than the age of the cluster, it is flagged as disrupted by dynamical friction.

The MOSAICS model for cluster formation and evolution has been used in a number
of studies that have successfully reproduced many observed properties of GC
populations.  The cluster disruption used in MOSAICS has been shown to reproduce
observed $z=0$ cluster distributions in age, space, and mass, as well as the
kinematics of GC systems
\citep{Kruijssen2011,Kruijssen2012b,Adamo2015,Miholics2017}. The simulated
E-MOSAICS galaxies and GC populations reproduce the typical star formation
history, specific frequency, metallicity distribution, mass function, spatial
distribution \citep{Pfeffer2018,Kruijssen2019a} and ``blue tilt''
\citep{Usher2018} seen in local $L^*$ galaxies.  The young stellar clusters in
E-MOSAICS reproduce cluster formation efficiency and high mass truncation in the
initial cluster mass function \citep{Pfeffer2019b} observed in the local galaxy
population.

The E-MOSAICS simulations have also been used to study a broad variety of
questions in cluster and galaxy formation and evolution.  They have been used to
connect the galactic assembly history to the GC population in age-metallicity
space \citep{Hughes2019,Kruijssen2019a,Kruijssen2019c} and to the $\alpha$ element
abundances of GCs \citep{Hughes2020}.  E-MOSAICS has been used to predict the
GC UV luminosity function \citep{Pfeffer2019a}, the overall formation history of
clusters and GCs compared to field stars \citep{Reina-Campos2019}, and constrain
the fraction of halo stars contributed by dynamical disruption of GCs
\citep{Reina-Campos2020}.

\section{The birth environment of globular clusters}
\label{s:birthenv}
\begin{figure*}
    \includegraphics[width=\textwidth,keepaspectratio]{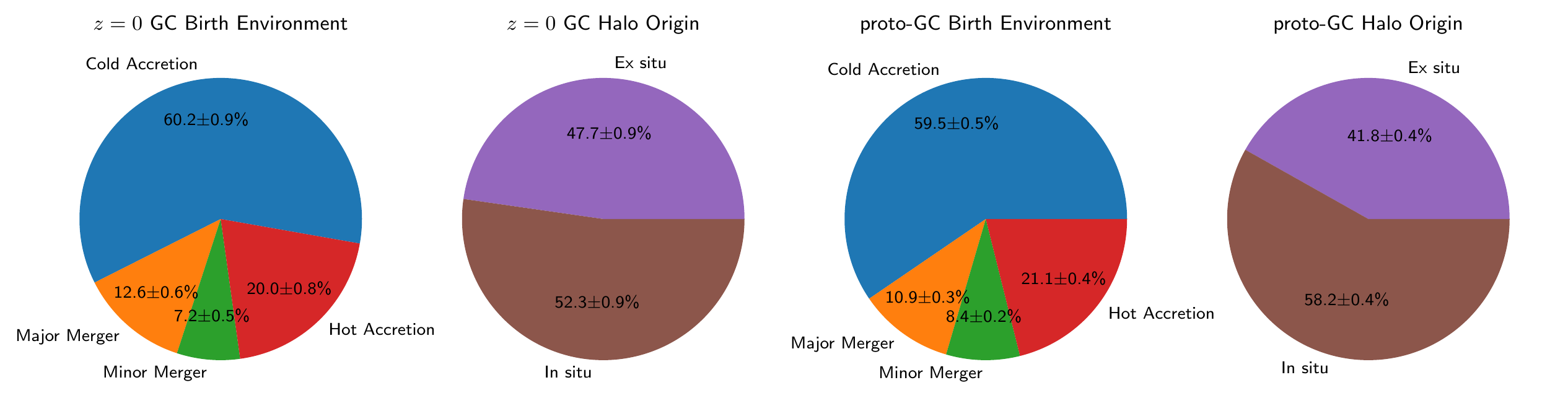}
    \caption{The birth environment and natal galaxy of both the present day GCs
    and all formed proto-GCs. For both the surviving GC population and the
    proto-GCs primarily form out of gas which has never been shocked above
    $2\times10^5\K$, with $15-20$ per cent forming during major and minor
    mergers.  Most proto-GCs and GCs at $z=0$ were formed in-situ ($58.2 \pm
    0.7$ per cent and $52 \pm 1.0$ per cent, respectively).  The larger fraction
    of in-situ proto-GCs indicates that proto-GCs formed in-situ experience
    stronger disruption over their lifetimes than those formed in satellites
    that were subsequently accreted.}
    \label{pie_charts}
\end{figure*}
\begin{figure*}
    \includegraphics[width=\textwidth]{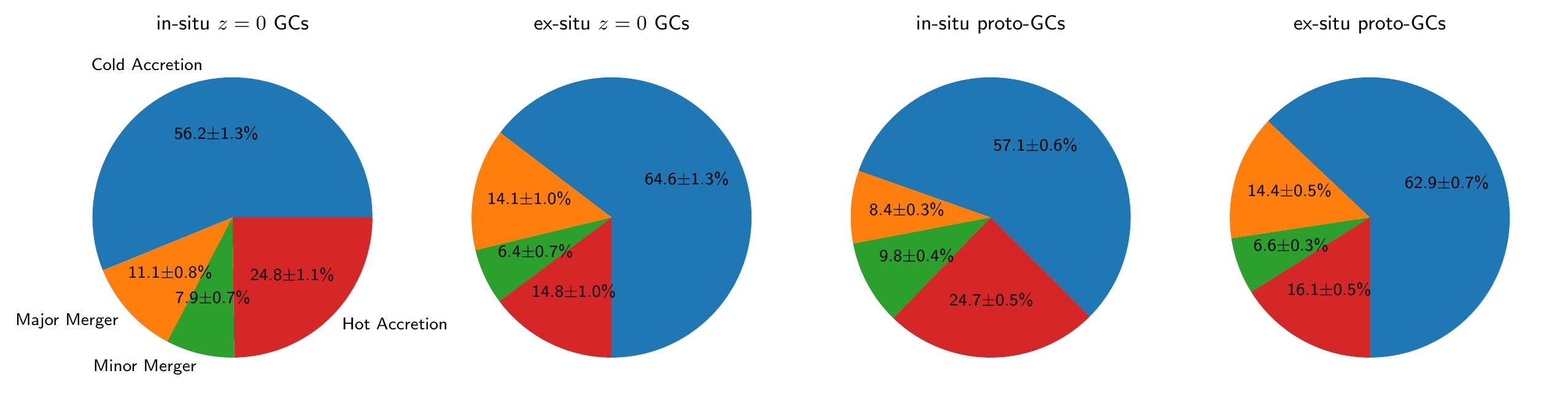}
    \caption{The birth environment for in-situ and ex-situ $z=0$ GCs and
    proto-GCs.  In-situ clusters form within the ``trunk'' of the merger tree,
    while ex-situ clusters form in smaller haloes that are then accreted.  The
    ex-situ population contains a roughly equal fraction of merger-formed GCs
    ($20.5 \pm 1.2$ per cent as opposed to $19.1 \pm 1.1$ per cent) and a lower fraction
    of GCs forming out of shock-heated gas ($15 \pm 1$ per cent as opposed to
    $24.7 \pm 1.1$ per cent).  This is to be expected, as lower-mass haloes have
    lower virial temperatures.}
    \label{pie_charts2}
\end{figure*}
\begin{figure*}
    \includegraphics[width=\textwidth]{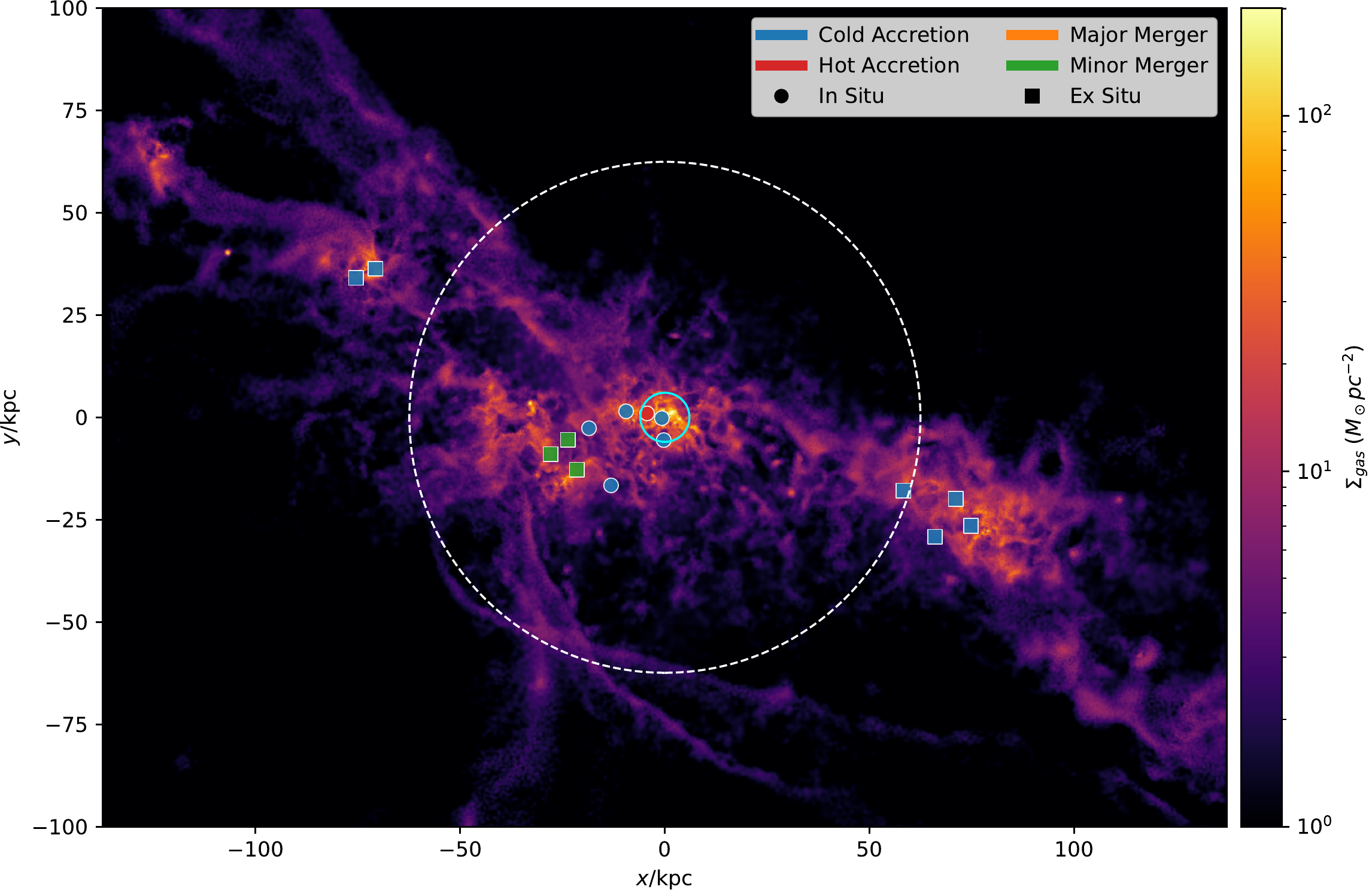}
    \caption{Gas column density in the environment of a typical GC-forming
    galaxy at $z=2.5$.  Gas particles that will form GCs by the next snapshot
    are marked with coloured circles (in-situ) or squares (ex-situ), each
    indicating their formation channel (using the same colour scheme as in
    Figure~\ref{pie_charts} and~\ref{pie_charts2}).  The virial radius of the
    galaxy is shown with the dashed white circle, and twice the stellar
    half-mass radius is shown with the solid cyan circle.  The majority of the
    GCs that form at this time come from the cold, clumpy medium fed by the many
    visible filaments.  A minor merger in the outskirts of the disc will produce
    3 ex-situ GCs, and a single cluster is formed from gas which once exceeded a
    temperature of $2\times10^{5}\K$.  As can be seen, two smaller haloes will
    deliver 6 ex-situ clusters to the central galaxy when they later merge.
    These mergers will occur {\it after} these GCs form in the satellite
    galaxies.}
    \label{example_image}
\end{figure*}
We select $z=0$ GCs using the same criterion as \citet{Kruijssen2019a}, namely
that their masses must be $>10^5\Msun$, metallicities $\rm [Fe/H]$ between
$-2.5$ and $-0.5$, and with galactocentric radii of $>3\kpc$, to select clusters
that match observed GC properties and exclude clusters that likely have
experienced underdisruption due to numerical effects (e.g. strongly softened
gravity and importantly, under-resolved cold ISM clumps).  With these selection
criteria, we find 2732 GCs across the 25 E-MOSAICS galaxies.  We also apply the
same selection criteria, but rather than using the $z=0$ masses of the clusters,
instead use their birth masses.  This allows us to generate a population of
``proto-GCs'', of clusters that would be identified as GCs at $z=0$ had they not
experienced disruption due to their individual tidal histories (or lost mass due
to stellar evolution).  We find 12,186 of these proto-GCs, showing once again
(as has been reported in \citealt{Reina-Campos2018}) that dynamical disruption
plays as much of a role producing the present-day GC population as does the
formation physics.  For all quantities related to population fractions, we
calculate a mean estimate and confidence interval using 1000 bootstrap samples
\citep{Efron1979}.

\subsection{Distribution of GCs across different formation channels}
We split the populations of GCs and proto-GCs into a set of disjoint subsets
(the union of which bijects the whole population).  The first family is the
birth environment, and contains four possible formation channels.  A cluster can
form during a merger event (either a major merger, with a stellar mass ratio
above 1:4, or a minor merger with a stellar mass ratio between 1:4 and 1:10).
We identify merger events by selecting subhaloes which formed through the merger
of two or more subhaloes from the previous snapshot, and identify clusters which
formed in that interval as forming during a merger.  If a cluster does not form
during a merger, it may form from gas that has never been strongly shock-heated
above the $2\times10^5\K$ peak of the cooling curve (gas that we identify as
coming from ``cold accretion''), or gas which has been heated above this
temperature (gas we identify as brought to the galaxy by ``hot accretion'').  As
this is near the peak of the cooling curve (due to efficient recombination
cooling), gas will only be pushed above this temperature by feedback or a virial
shock, and will usually persist well above or below this temperature
\citep{Brooks2009,Woods2014}.  This is also roughly the virial temperature that
galaxies will reach once their halo mass exceeds $10^{11}\Msun$.  As we simply
track the maximum temperature a gas particle reaches prior to star formation,
without a detailed analysis of the full thermal history, this temperature split
will not perfectly map to gas accreted along cold flows vs.  gas shock-heated at
the virial radius.  Heating from SNe can also generate temperatures above
$2\times10^5K$, and failure to capture brief shocks may miss brief excursions
above $2\times10^5\K$.  Rather than specifically referring to gas accreted along
cold streams as opposed to gas heated by a virial shock, our ``cold accretion''
and ``hot accretion'' populations are more accurately described simply as gas
which was always cold as opposed to gas that was previously heated.  Naturally,
since star formation in the EAGLE model occurs when gas cools below the $8000\K$
equation-of-state threshold, gas that will form ``hot accretion'' GCs must cool
before star formation actually occurs. We also split the population of GCs and
proto-GCs based on whether they form in the largest progenitor of the $z=0$
galaxy (``in-situ'' clusters, in the trunk of the merger tree), or in one of the
smaller progenitors (``ex-situ'' clusters, formed in the branches of the merger
tree).

In Figure~\ref{formation_environment}, we show the number of clusters formed
through each of these four channels for the 25 E-MOSAICS MW analogues.   The
majority of GCs in each galaxy (except for MW10) form outside of merger events,
with most of these forming from gas which has never been heated above
$2\times10^5\K$.  It is also clear here that for MW-mass galaxies, there is
fairly significant scatter of both the fraction of GCs formed through each
channel, and the total number of GCs formed (this has been previously shown
explicitly in figure 2 of \citet{Kruijssen2019a}, and is implied by studies of
the specific frequency of GCs, such as \citealt{Peng2008}).  As
Figure~\ref{pie_charts} shows,  both the $z=0$ GC population, as well as the
proto-GCs form primarily outside of merger events ($>80$ per cent of each population),
and mostly from gas accreted onto their parent galaxy without being heated.
Galaxy mergers appear to be a fairly minor contributor to the GC population,
accounting for $15-20$ per cent of the proto-clusters that are formed, as well
as the fraction of those which survive to $z=0$.

In Figure~\ref{pie_charts}, we also show the fractions of GCs and proto-GCs that
form in-situ or ex-situ.  The fraction of proto-GCs formed in-situ is somewhat
larger than the fraction of in-situ GCs that survive to $z=0$ ($58.2 \pm 0.7$ per cent
vs. $52.3 \pm 1.3$ per cent).  This tells us that the disruption mechanisms that destroy
$\sim80$ per cent of the proto-GCs before $z=0$ are more effective for in-situ
clusters than ex-situ clusters.  A deeper potential well, as well as higher ISM
densities in the larger primary progenitor may help explain these differences,
because of the increased rate of cluster disruption and tidal shocks.

We can also separate the in-situ and ex-situ populations based on the four
formation channels previously examined.  In Figure~\ref{pie_charts2}, we show
the fraction of each formation channel producing the in-situ and ex-situ GCs and
proto-GCs.  Once again, there is little difference between the breakdown of
formation channels between the surviving $z=0$ GCs and the proto-GCs.  We do,
however, see a noticeable difference between the in-situ and ex-situ
populations.  Ex-situ clusters are more formed in major mergers ($14.1 \pm 0.9$ per
cent in ex-situ, as opposed to $11.1 \pm 0.9$ per cent for in-situ), and less often
from gas that has been shock-heated ($14.8 \pm 1.0$ per cent for ex-situ as opposed to
$24.7 \pm 1.1$ per cent for in-situ).  We verify that this is not simply an effect of
the different median formation times by looking at only clusters formed before
$z=1$, and find similar fractions. The lower fraction formed by shock-heated
gas is easily understood: the less-massive haloes in which these clusters form
had virial temperatures below $2\times10^5\K$, and the stellar feedback that
could alternatively heated gas was less intense in the less active star
formation environment of these lower-mass haloes.  The larger fraction being formed
by mergers is interesting, as it suggests that these smaller, accreted galaxies
were less likely to produce the high ISM pressure and density required to form
GCs outside of merger events.

An example of the qualitative morphology of a typical galaxy in which the
majority of GCs form is shown in Figure~\ref{example_image}, which shows the gas
surface density in the progenitor halo of MW00 at $z\sim2.5$.  The bulk of the
clusters formed between this snapshot and the next are formed in-situ, from cold
accreted gas.  A handful of ex-situ clusters will also be delivered to the
central galaxy, from the two satellites falling in along the two primary
filaments.  These clusters will form in the two satellites prior to their
merging with the central halo, and form from gas that has never been shock
heated.  Dense gas in this galaxy clearly extends well past the stellar
half-mass radius, and is in a turbulent, highly disrupted state.  The single
example of a cluster formed from shock-heated gas is a case where the heating
comes not from a virial shock, but from stellar feedback, as the virial
temperatures of all haloes in this snapshot are $<2\times10^5\K$. 

\subsection{Properties of globular clusters formed through different channels}
\begin{figure}
    \includegraphics[width=\columnwidth]{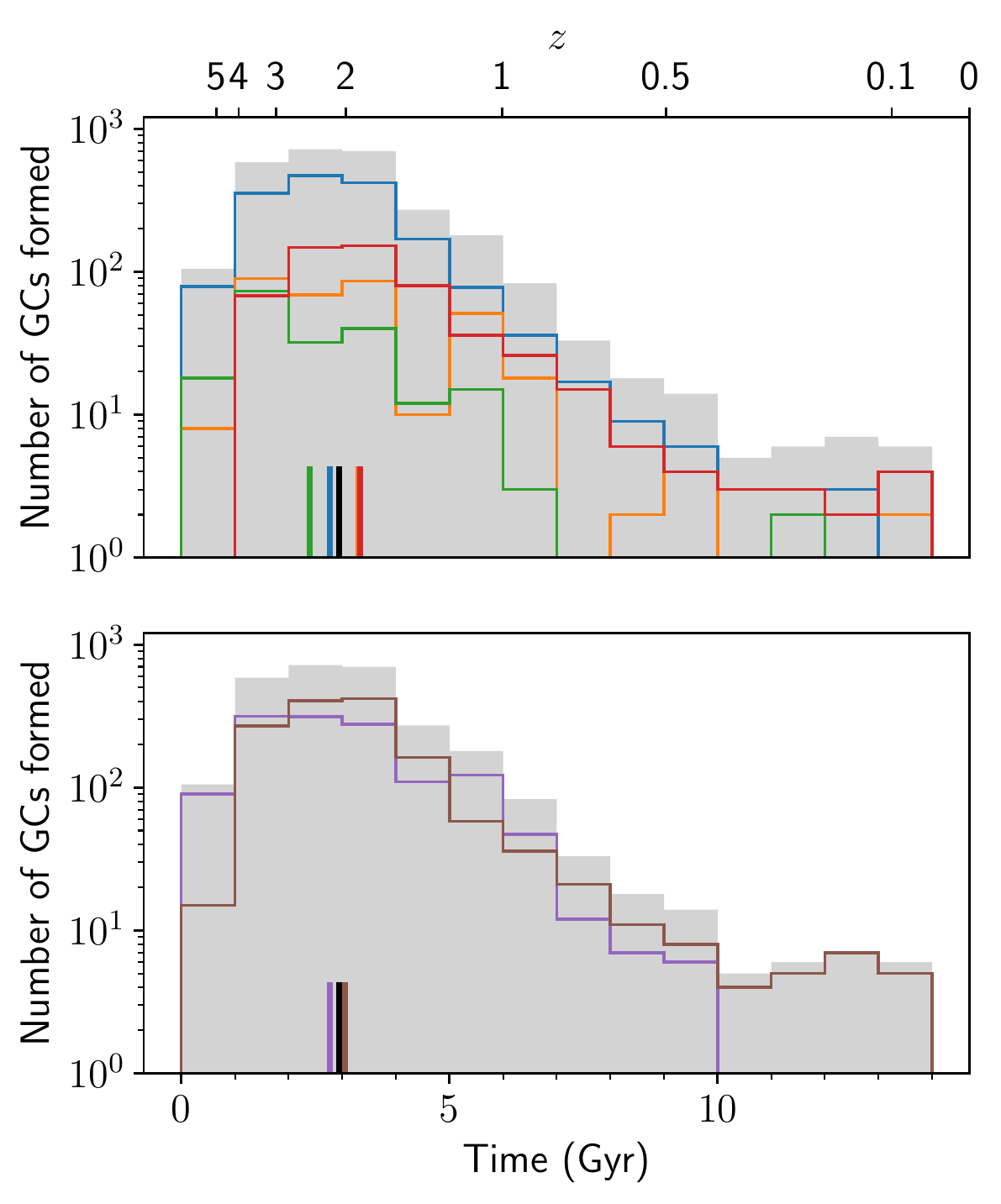}
    \caption{Formation times for $z=0$ GCs as a function of their formation
    channel (top) and whether the clusters formed in-situ vs. ex-situ (bottom).
    The top histogram shows the formation times for all GCs (grey), those formed
    in major mergers (orange), those formed in minor mergers (green), those
    formed by gas that was never shock-heated above $2\times10^5\K$ (blue), and
    those formed by gas that was shock-heated (red).  The vertical lines show
    the median formation times for each channel.  The typical GC
    forms at $z=2$, roughly $11.3\Gyr$ ago.  The bottom histogram shows the same
    quantities, but split based on whether the cluster forms in-situ (brown) or
    ex-situ (purple).  Here it can be seen that the formation rates as a
    function of time for both in-situ and ex-situ clusters are relatively
    similar.}
    \label{redshift_split}
\end{figure}
\begin{figure}
    \includegraphics[width=\columnwidth]{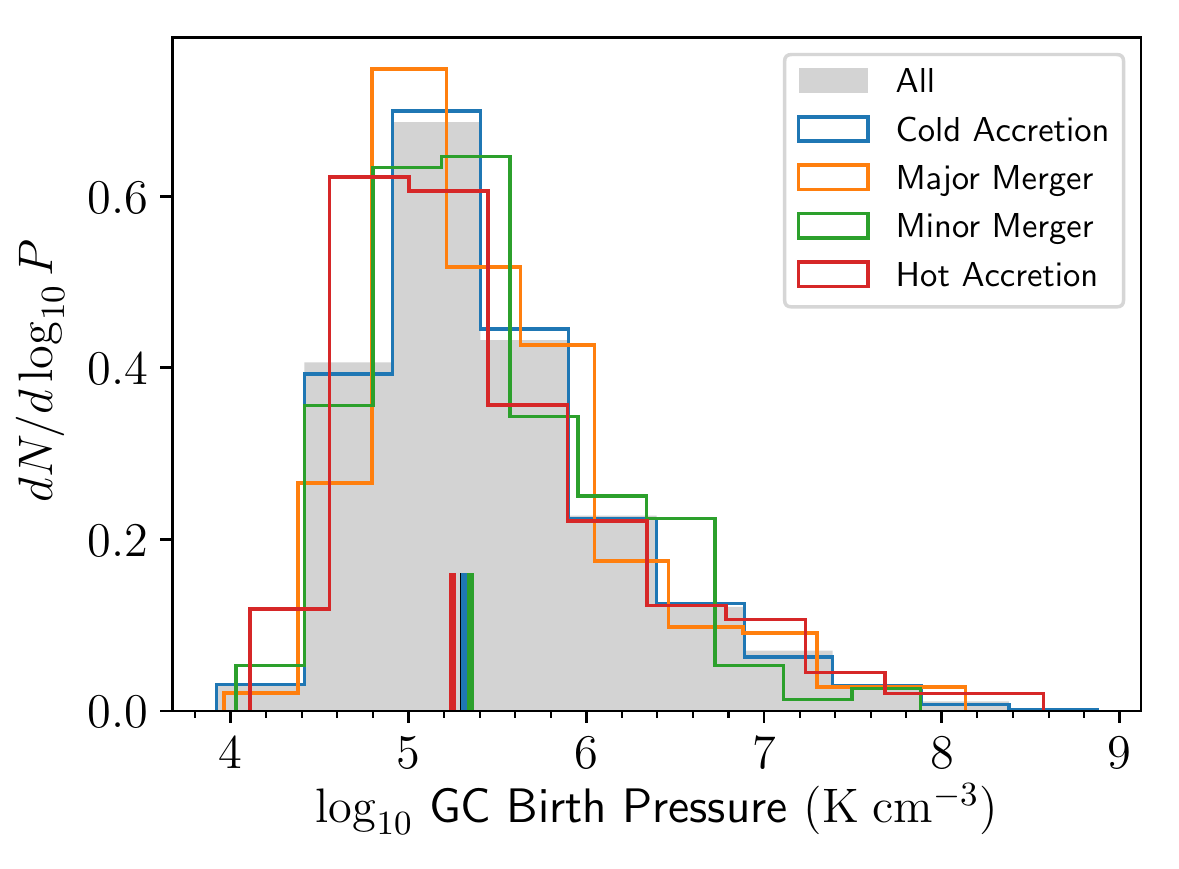}
    \caption{Distribution of ISM birth pressure for GCs as a function of their
    formation channel.  The distribution of birth pressures for all
    GCs (grey), those formed in major mergers (orange), those formed in minor
    mergers (green), those formed by gas that was never shock-heated (blue), and
    those formed by gas that was shock-heated (red) are effectively
    indistinguishable.  Regardless of what the state of the galaxy is, or how
    the gas which forms the GC is accreted onto the galaxy, the distribution of
    birth pressures is effectively the same, with the average GC forming from
    gas with $P/k_{\rm B} = 10^{5.3\pm0.9} \K\hcc$.}
    \label{pressure_split}
\end{figure}
\begin{figure}
    \includegraphics[width=\columnwidth]{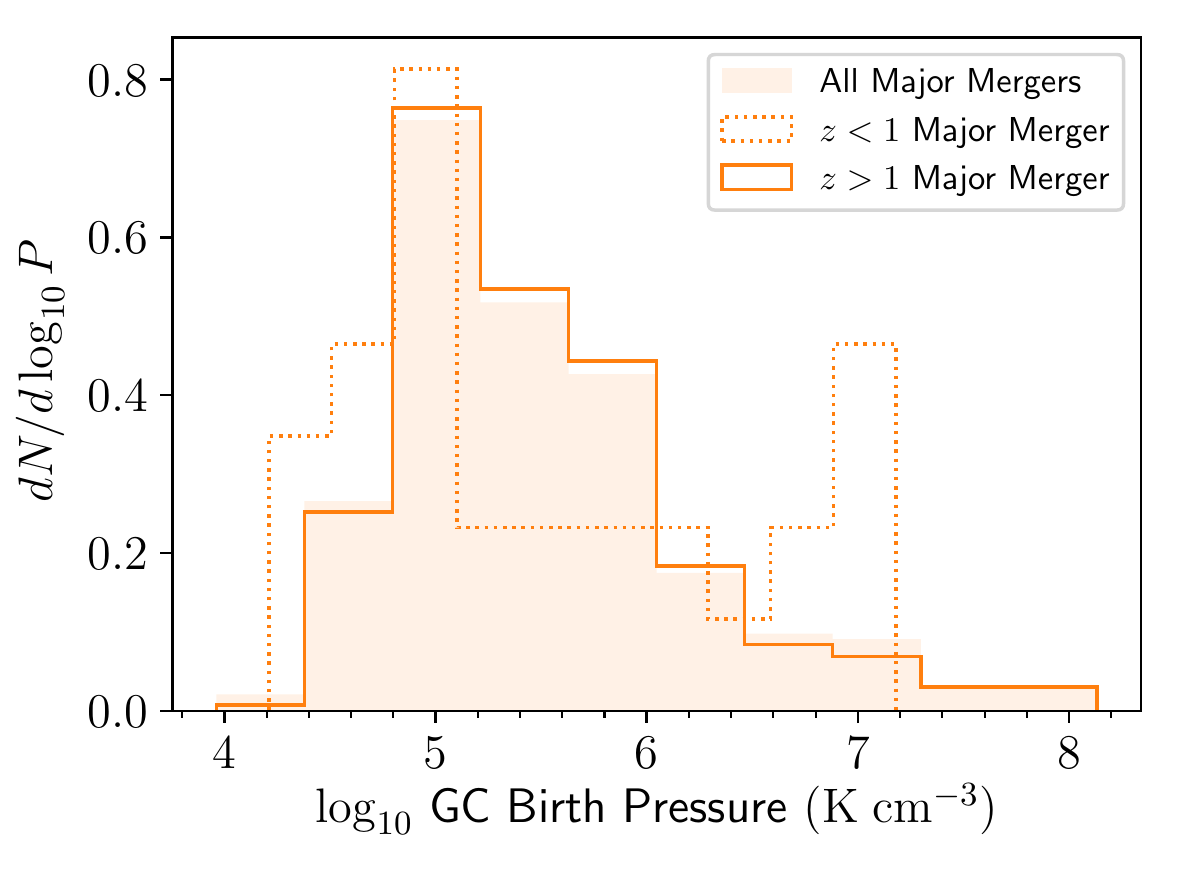}
    \caption{Distribution of ISM birth pressure for GCs formed in major mergers.
    The solid histogram shows all GCs formed in major mergers, the dashed line
    shows GCs formed in major mergers after $z=1$, and the solid line shows GCs
    formed in major mergers before $z=1$.  The old GCs make up the majority of
    the population that formed in major mergers, and thus track the average
    pressure distribution.  Notably, the young major merger-born GCs show a
    distinct bimodal distribution, with a high-pressure population forming from
    gas with pressures of $P/k_{\rm B}\sim10^7\K\hcc$.}
    \label{mergers_pressure}
\end{figure}
\begin{figure}
    \includegraphics[width=\columnwidth]{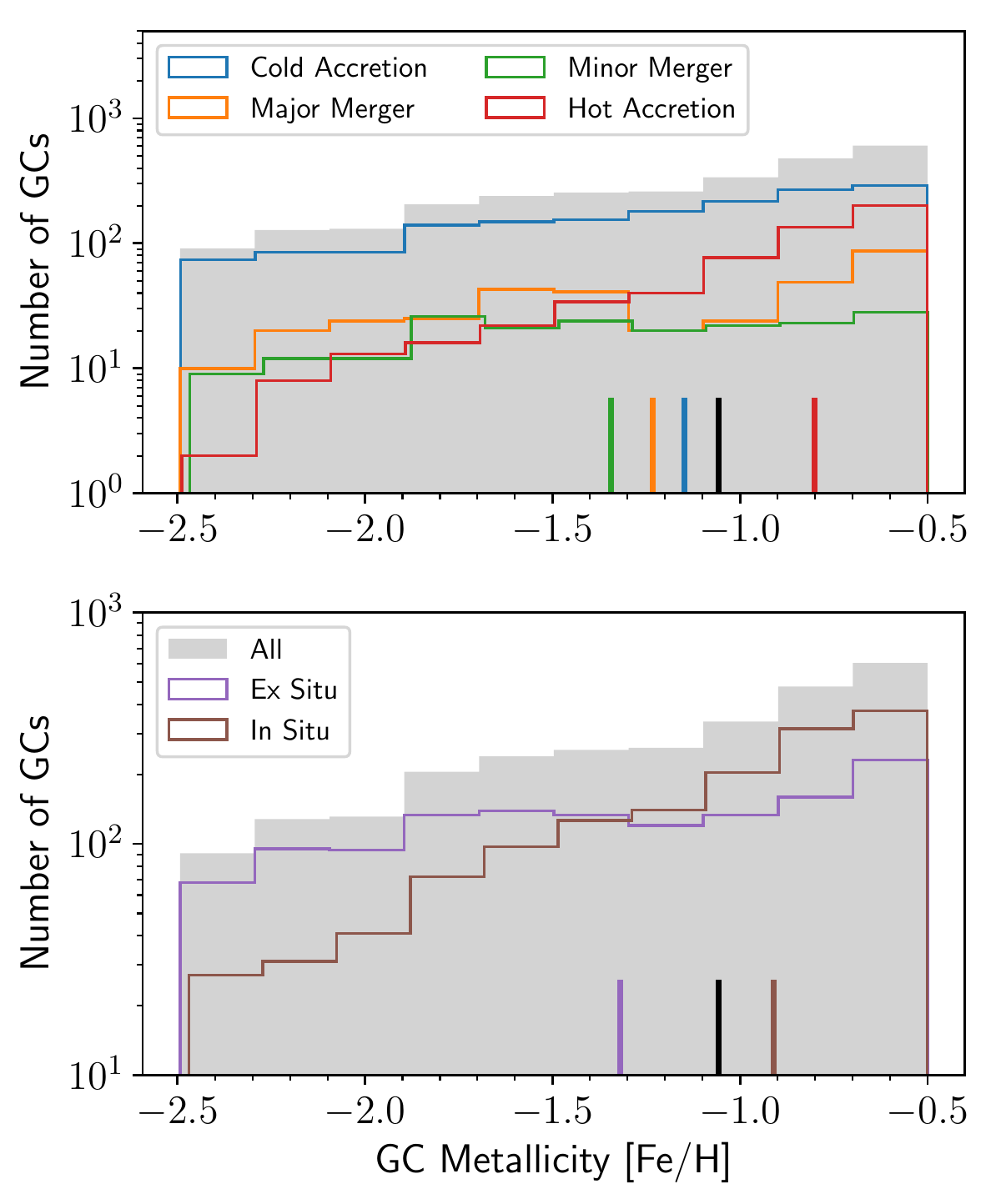}
    \caption{Metallicity distributions of GCs as a function of formation channel 
    (top) and in-situ or ex-situ formation (bottom).  As the top
    panel shows, GCs formed through mergers have a much flatter metallicity
    distribution than those formed through smooth accretion (whether that
    accretion is cold or hot).  Clusters formed through hot accretion have the
    steepest metallicity distribution, with the vast majority having high
    metallicities.  GCs formed during major mergers show a nearly uniform
    metallicity distribution.  In the bottom panel, we see that the clusters
    formed ex-situ have significantly lower metallicities than those formed
    in-situ, as well as having a much flatter metallicity distribution.
    Clusters with metallicity $\rm [Fe/H] < -1.5$ are dominated by the ex-situ
    component.}
    \label{metallicity_split}
\end{figure}
\begin{figure}
    \includegraphics[width=\columnwidth]{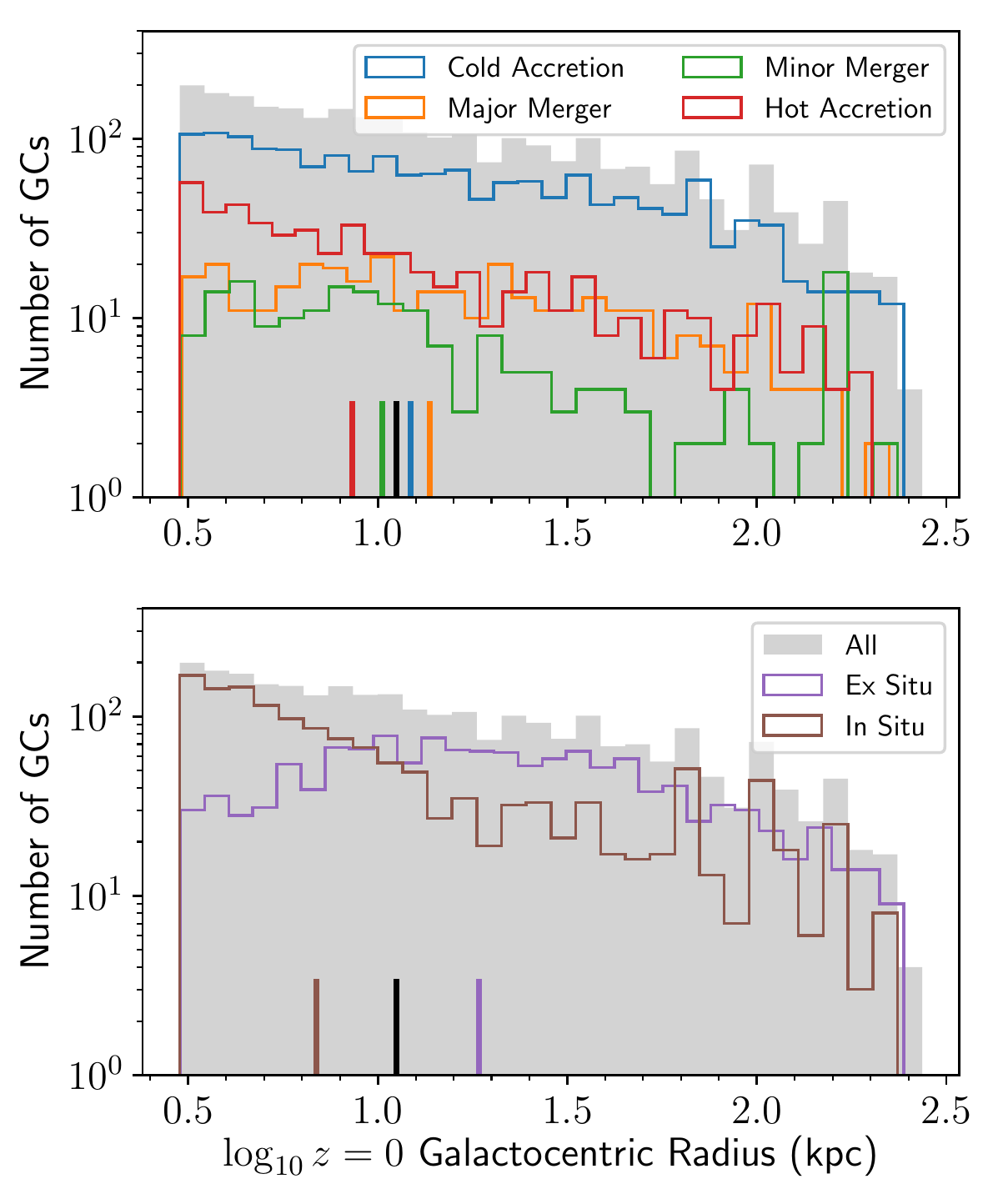}
    \caption{Galactocentric radii of GCs as a function of formation channel (top
    panel) and in-situ vs. ex-situ formation (bottom).  As the
    top panel shows, GCs formed through smooth accretion (red and blue curves) have a
    more centrally peaked radial distribution, with the median cluster falling at
    galactocentric radii of $\sim10\kpc$, while those formed through mergers
    (orange and green curves) fall into a nearly uniform radial distribution.
    In the bottom panel, we see this is even more pronounced when we
    differentiate between in-situ and ex-situ clusters.}
    \label{radius_split}
\end{figure}
The natural question these channels raise is whether populations of GCs formed
in each channel share properties that are distinguishable from populations
formed through the other channels, and whether these differences are potentially
observable.  Looking simply at when the present-day clusters are formed through
each channel, as we do in Figure~\ref{redshift_split}, we see that most GCs are
formed between $z=1.5-5$, when the Universe was less than one third of its current
age (also see \citealt{Reina-Campos2019}).  We can also see noticeable transitions between
the frequency of GCs formed through each of the four channels we examine.  While
the majority of GCs form from unshocked gas outside of merger events at nearly
every epoch, the youngest population of clusters (especially those formed after
$z=1$) is increasingly composed of those formed during major mergers and from
shock-heated gas (prior to $z=1$, $>60$ per cent of clusters form from cold
accretion, versus $\sim 50$ per cent after $z=1$).  The oldest population is those
formed in minor mergers, with a median birth time of $2.39\Gyr$ after the Big
Bang, while the youngest is those formed through major mergers, with a median
formation time of $3.34\Gyr$.  GCs formed from previously heated gas have a
median formation time of $3.31\Gyr$, and those formed from unheated gas have a
median formation time of $2.77\Gyr$.  The median formation time for all GCs is
$2.93\Gyr$.  We would expect the increase in shock-heating as the haloes become
more massive and their virial temperatures increase, as well as 
more opportunities for feedback heating as the ISM is recycled in galactic fountains.
Interestingly, the component formed from minor mergers is almost entirely formed
prior to $z=1$, making them (on average) the oldest population, along with those
which are formed from cold-accreted gas. For in-situ vs.  ex-situ clusters,
however, we see no difference in the median formation times, with both
populations roughly tracking the total formation rate, and with median formation
times comparable to each other.

If we look at the ISM property which primarily sets the cluster formation
efficiency ($\Gamma$), the pressure, we see in Figure~\ref{pressure_split}
little difference between clusters formed through each of our channels.   The
ISM birth pressures for GCs formed from each of the examined channels (as well
as the full population) have nearly identical distributions.  The median ISM
pressure which $z=0$ GCs are formed is $P/k_{\rm B} = 10^{5.3\pm0.9}
\K\hcc$.  This distribution has noticeable skewness, with a long tail of
clusters formed form pressures as high as $P/k_{\rm B}=10^9\K\hcc$.   We can also see
the enhanced pressures that are observed with YMC formation sites at
low redshift in Figure~\ref{mergers_pressure}.  At low redshift, major mergers
can form a larger fraction of GCs from much higher pressures than at earlier
times, producing a distinct population of GCs formed at pressures of
$P/k_{\rm B}\sim10^7\K\hcc$.  The higher collision velocities and more massive
galaxies involved in recent major mergers make these events more violent,
compressing the ISM to higher pressures (but for shorter time periods) than what
is typical at higher redshifts.

If we look in Figure~\ref{metallicity_split} at the metallicity of the clusters,
probed through their iron abundances, we see a few notable features of our
different formation channels. For all metallicity bins, GCs formed through cold
accretion make the largest fraction of GCs.  The metallicity distribution for
clusters formed through minor mergers is nearly flat, resulting in the lowest
median metallicity of $\rm [Fe/H]\sim-1.3$, while the metallicity distribution
for clusters formed after being shock-heated is strongly biased towards
metal-rich GCs, with a median $\rm [Fe/H]\sim-0.80$.  These medians are somewhat
unsurprising when we consider that minor mergers are frequent at high redshift
(when the cosmic metallicity was lower), and will tend to bring relatively
pristine gas in with the smaller merging partner (as the smaller partner,
following the mass-metallicity relation, will have lower overall metallicity,
see e.g. \citealt{Erb2006}).  For GCs formed from previously shock-heated gas,
this gas was either heated through a virial shock, implying the halo mass
$>10^{11}\Msun$, and thus a much more metal-enriched ISM (assuming that the
mass-metallicity relation holds), or through SN feedback, which brings with it
fresh metals in the form of the SN ejecta.  As GCs of all metallicities are
primarily formed outside of mergers, from unshocked gas, we cannot easily use a
simple metallicity criterion to determine observationally whether a cluster is
formed through a given channel.

We do, however, see that if we look at metallicity for in-situ or ex-situ
clusters, the median metallicity for in-situ clusters is larger ($\rm
[Fe/H]\sim-0.9$ as opposed to $\rm [Fe/H]\sim-1.3$), and the distribution is
much steeper.  While some individual E-MOSAICS galaxies show bimodal MW-like GC
metallicity distributions, we see here that neither the full population across
all 25 galaxies, nor the population of in-situ or ex-situ GCs is simply bimodal.
We do see that extremely metal poor GCs, with $\rm [Fe/H]<-1.5$ are dominated
by ex-situ GCs.  Despite this, if we look at metal rich vs. metal poor GCs, we
cannot simply split the population based on their origin as in-situ or ex-situ.
Both scenarios contribute significant numbers of GCs, and as
\citet{Kruijssen2019a} showed, the old and metal-poor GC populations show a roughly
equal contribution of in-situ and ex-situ GCs.  Generally, ex-situ GCs tend to
be lest metal rich than in-situ GCs as a consequence of the mass-metallicity
relation and the fact that, for a given formation time, in-situ GCs form in more
massive haloes than ex-situ ones (by definition).  The fact that in-situ clusters
have typically formed in more massive objects (as indicated by their higher
metallicity) helps to explain their more effective disruption: more massive
objects will have stronger tides, and a greater chance for close encounters with
the ISM of the disc.  When the GCs are young, and subject to the most intense
tidal shocks from their natal environment (the dense ISM), this deeper potential
will make it more difficult for them to be ejected from the disc.

Turning to the spatial distribution of GCs , we see in Figure~\ref{radius_split}
that there is a noticeable difference between GCs formed through different
channels, but again at all radii the population is dominated by clusters formed
through cold accretion.  The most centrally concentrated population are those
formed through hot accretion, as we might expect from their higher metallicity
and the metallicity gradient observed in both real GC systems and the E-MOSAICS
galaxies. Clusters formed through mergers have a broader distribution, with
GCs formed in a major merger lying at a median radius of $\sim12\kpc$. In-situ
clusters have a median galactocentric radius of $\sim7\kpc$, while ex-situ
clusters have median radii of $\sim18\kpc$.  Interestingly, we can also see that
a slim majority of GCs in the outer halo, beyond $10\kpc$, are accreted ex-situ
clusters.  We also see that in-situ clusters can be deposited up to
$\sim100\kpc$ from the galaxy.  

\subsection{The galaxies that form GCs}
\begin{figure}
    \includegraphics[width=\columnwidth]{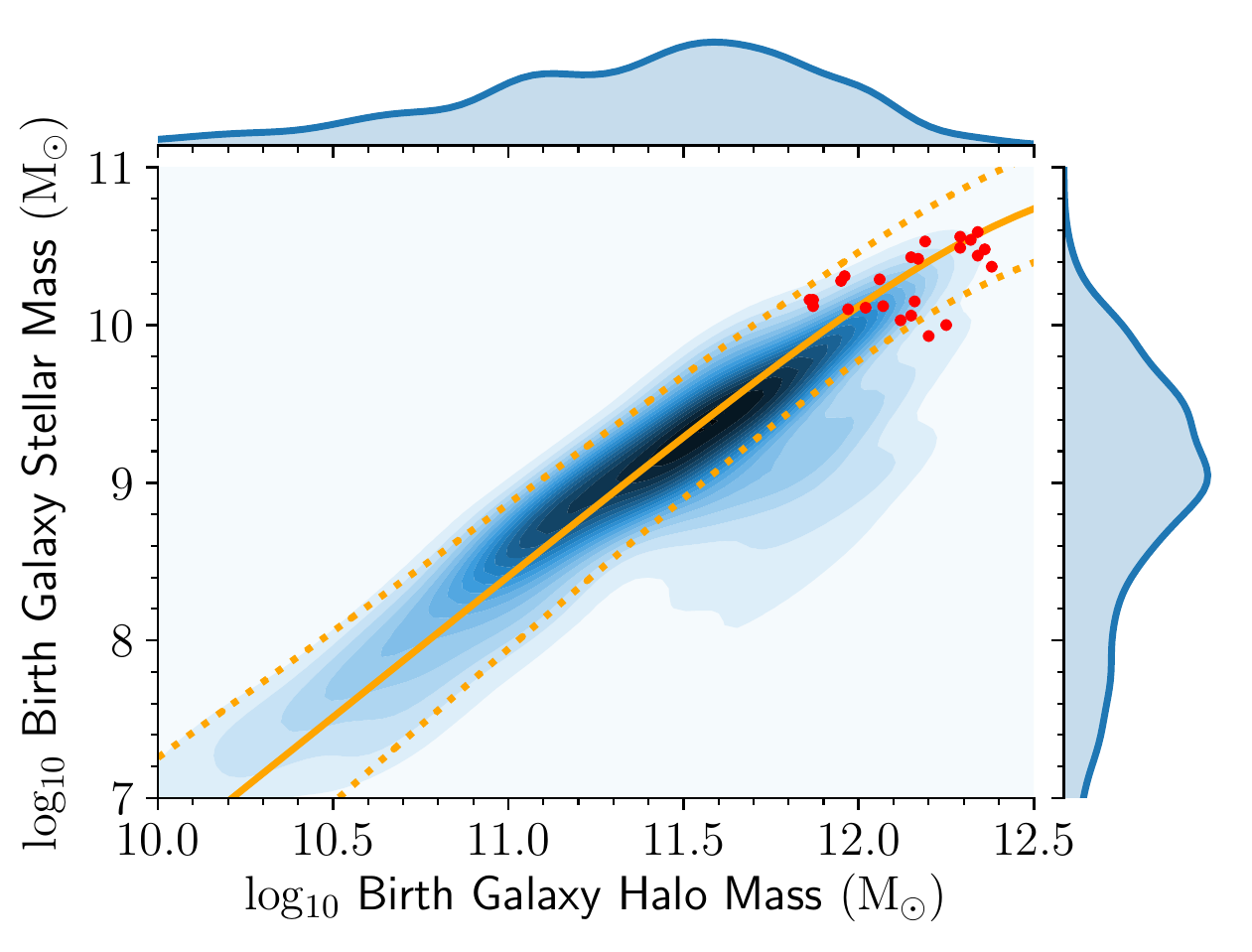}
    \caption{The relation between stellar mass and halo mass for haloes in which
    GCs are formed.  The main panel shows contours of the number of GCs formed,
    with Kernel Density Estimates (KDEs) shown for the halo and stellar mass
    above and to the right.  The median GC forms in a halo with virial mass
    $M_{200}=10^{11.5\pm0.7}$ and stellar mass $M_*=10^{9.0\pm1.1}$.  The orange
    lines show the \citet{Moster2013} SMHMR at $z=2$ (median in solid, scatter
    in dotted).  Red points show the final halo and stellar masses of the
    E-MOSAICS galaxies at $z=0$.  The vast majority of GCs form
    in relatively massive haloes (rather than low mass dwarfs), which lie on the
    typical SMHMR for the median GC formation redshift, $z\sim2$. The scatter
    seen in the E-MOSAICS haloes is  in part due to them forming at a variety of
    redshifts, while the overplotted curves are for haloes at a single redshift
    ($z=2$), but also due to the intrinsic scatter in the SMHMR.}
    \label{SMHMR}
\end{figure}
\begin{figure}
    \includegraphics[width=\columnwidth]{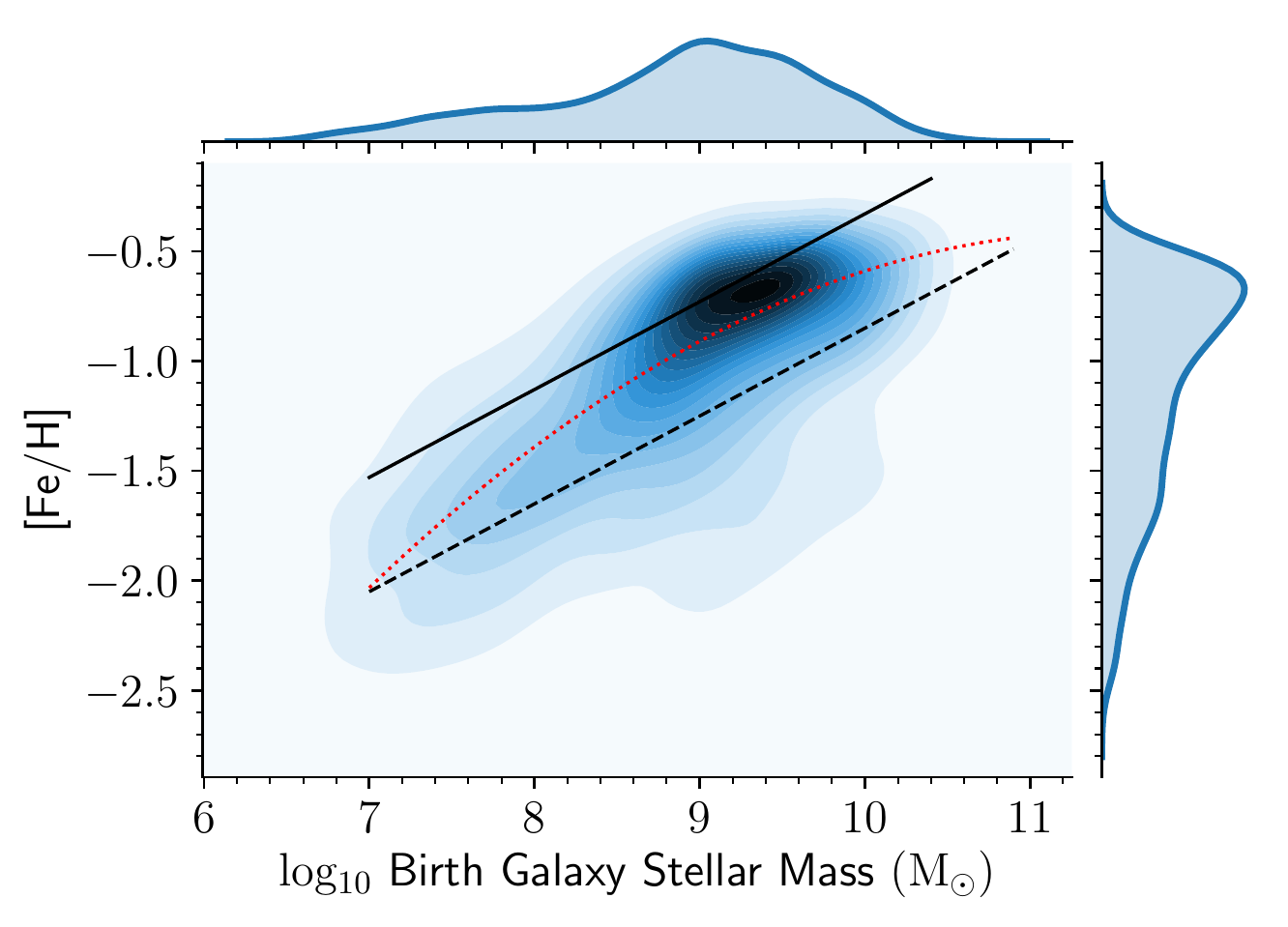}
    \caption{Globular cluster metallicity versus stellar mass of its birth
    galaxy.  There is a close relation between the metallicity of $z=0$ GCs and
    the stellar (and therefore halo) mass of the galaxy they form in, an
    indication that GCs can ``lock in'' the galaxy mass-metallicity relation,
    indicating directly the mass of the galaxy they formed in through their
    present metallicity.  The solid and dashed curves show the z$=0$ and $z=3$
    galaxy mass-metallicity relations from the \citet{Ma2016} simulations, while
    the red dotted curve shows the mass-metallicity relation at $z=2$ derived in
    \citet{Kruijssen2014b} from observations by \citet{Erb2006} and
    \citet{Mannucci2009}.}
    \label{MassMetal}
\end{figure}
As we have seen, most globular clusters are formed in-situ, in between periods
of major or minor mergers.  Figure~\ref{SMHMR} shows us that the galaxies GCs
are born within fall well on the abundance-matched host stellar mass--halo mass
relation (SMHMR) of \citet{Moster2013} for the typical redshift at which most of
our GCs form ($z\sim2$).  There is a relatively broad distribution in both the
stellar and halo masses of the galaxies in which GCs form, with typical virial
masses of $M_{200}=10^{11.5\pm0.7}\Msun$ and stellar mass
$M_*=10^{9.0\pm1.1}\Msun$.

We have already seen that there are some noticeable differences in the
metallicity distribution of GCs formed through different channels, and whether
those clusters form in-situ or ex-situ.  In Figure~\ref{MassMetal}, we show that
the metallicity of $z=0$ GCs trace, with significant scatter, the stellar mass
of their birth galaxy \citep[also see][]{Kruijssen2019a,Kruijssen2019c}. The
metallicity of GCs is slightly above the average ISM metallicity we would expect
from the mass-metallicity relation at $z\sim2$, when most GCs form.  This is
indicative of correlated star formation producing GCs: i.e.\ they form within
regions where previous star formation activity has enriched their natal gas with
metals.  The birth galaxy mass-GC metallicity relation explains most of the
differences between GCs of different origins seen in
Figure~\ref{metallicity_split}.  Clusters which are formed in accreted
satellites, or at higher redshift in minor mergers, will have lower metallicity
if their birth galaxy follows the mass metallicity relation (as those birth
galaxies will be less massive).

\begin{figure}
    \includegraphics[width=\columnwidth]{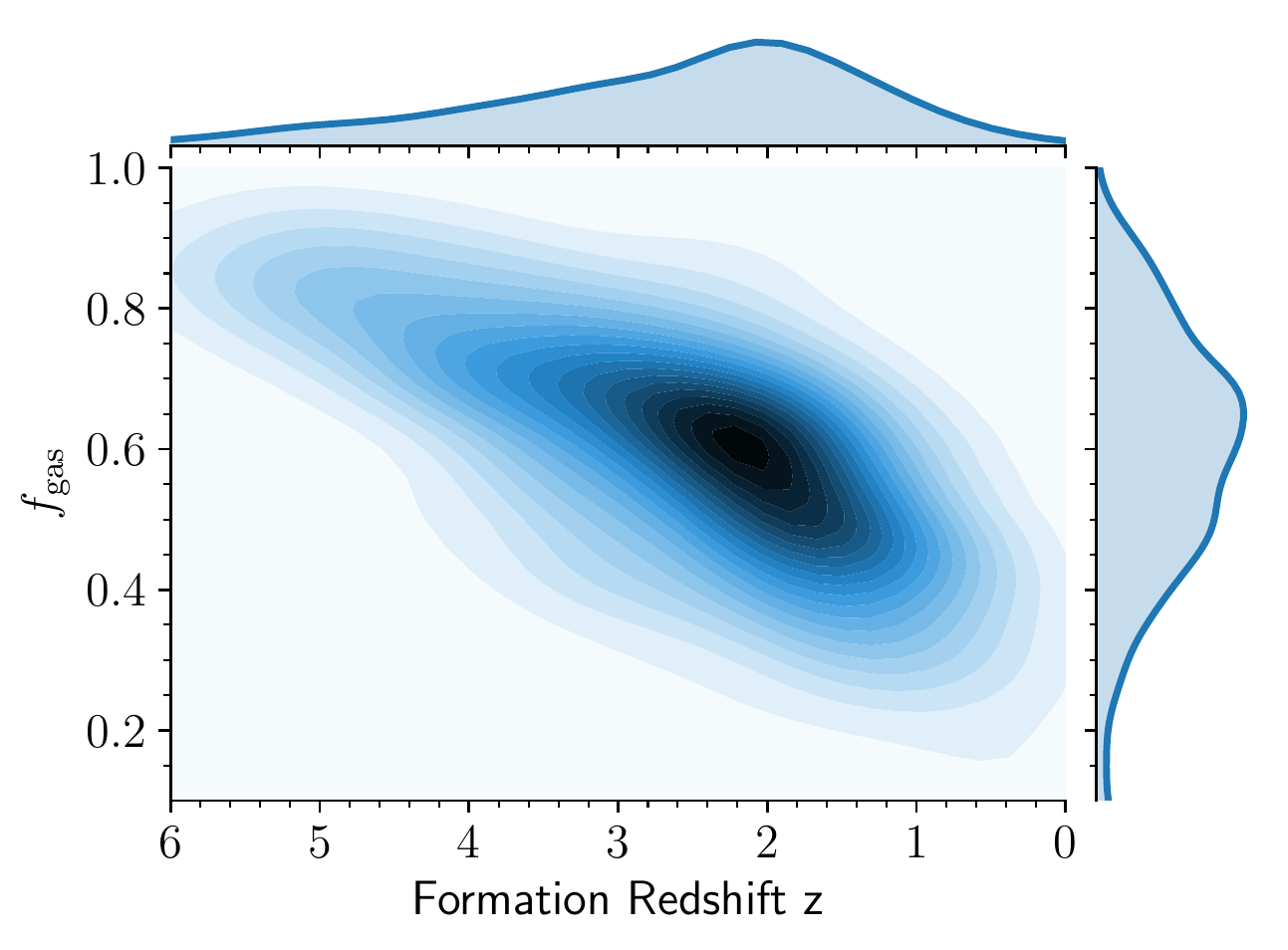}
    \caption{Subhalo gas fraction as a function of formation redshift of GCs.
    The majority of GCs form in highly gas-rich discs $f_{gas} > 0.5$,
    regardless of redshift.  However, a small fraction of GCs forms out of more
    stellar-dominated discs, predominantly at later redshifts, once sufficient
    time has passed to build up a stellar disc.}
    \label{GasFrac}
\end{figure}

The local ISM that forms GCs naturally must be able to produce the pressures we
see in Figure~\ref{pressure_split}.  In Figure~\ref{GasFrac} we look at the cold
gas fraction $f_{\rm gas}=M_{\rm cold}/(M_{\rm cold}+M_*)$ in the subhaloes
where GCs form.  Cold gas in this case is identified as gas below the
temperature threshold for star formation in E-MOSAICS, $2.5\times10^4\K$.  The
majority of GCs form in extremely gas-rich discs ($f_{\rm gas}=0.6\pm0.2$).  We
see also a decreasing trend in gas fraction as we move to lower redshift, as
shown by observations of high-redshift, star-forming galaxies
\citep{Tacconi2013}.  These gas fractions may be larger than those
observationally determined, as we are measuring the total cold gas content of
the subhalo, rather than the observationally-detected cold gas (traced by HI and
CO, where the latter dominates at these gas pressures) within the stellar disc.
This means we will include cold gas within the circumgalactic medium that might
otherwise be missed by looking only within the disc of the galaxy, though this
is likely a small minority.

\begin{figure}
    \includegraphics[width=\columnwidth]{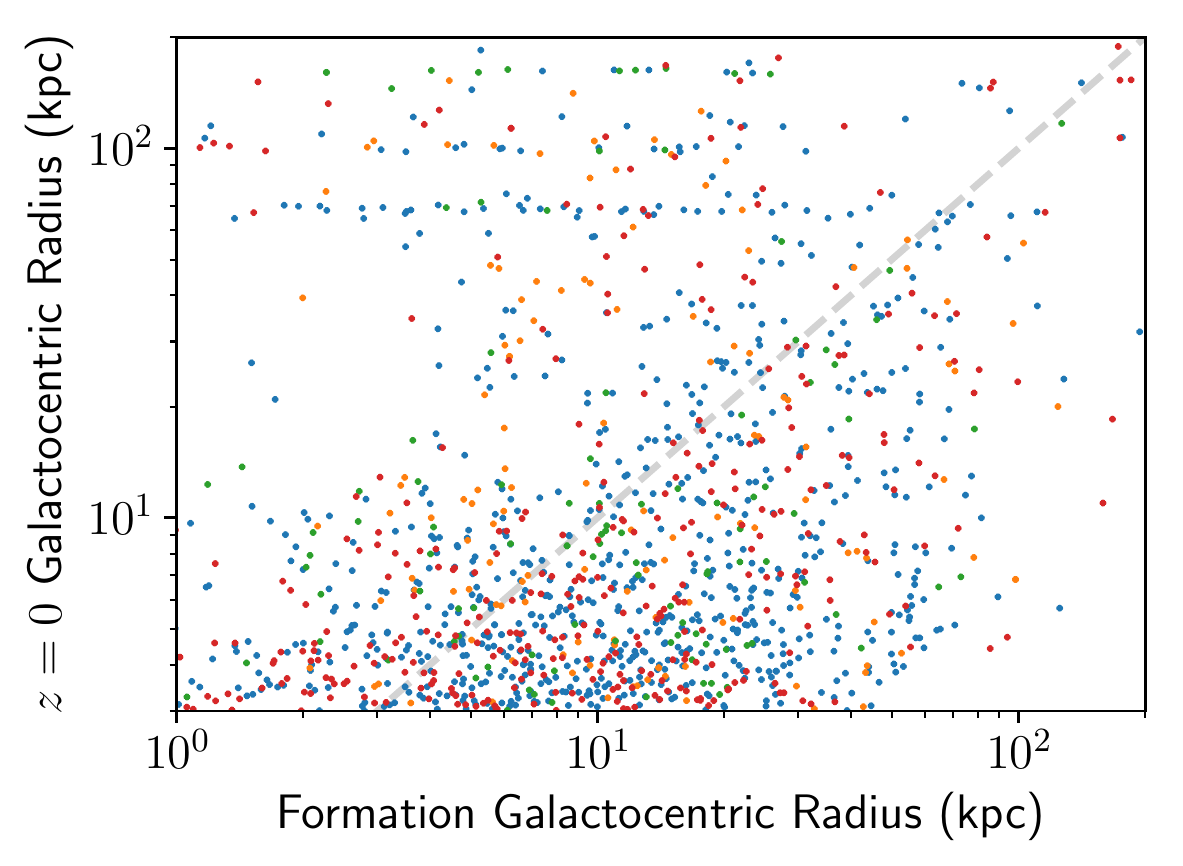}
    \caption{Final position at $z=0$ as a function of formation position for all
    in-situ GCs in E-MOSAICS, coloured by formation mechanism.  The light grey
    line shows the 1-to-1 line we would expect if GCs remained on the same
    orbital radii as they formed.  There is a tremendous amount of
    re-shuffling for all of the different formation mechanisms, with no obvious
    difference between them.  It is also clear that the majority of clusters end
    up on closer orbits at $z=0$ than they are born on.  This is somewhat to be
    expected from the evolution of the galaxy, as a GC with constant orbital
    energy will find itself in a tighter orbit as the halo mass grows. The
    horizontal lines seen near $\sim100\kpc$ are dwarf satellites that are within
    the primary halo's virial radius.}
    \label{radius_shuffle}
\end{figure}

For in-situ GCs, we can now look at how the assembly history of the galaxy
redistributes them onto new orbits.  In Figure~\ref{radius_shuffle}, we can see
how this radial shuffling takes place, comparing the galactocentric radius at
formation to the galactocentric radius at $z=0$.  Naturally, if GCs  have highly
eccentric orbits, we will find them at different galactocentric radii at
different times.  GCs on eccentric orbits spend more time at large radii, so we
would expect to find most GCs at larger radii than their formation radius.  On
the other hand, GCs forming in growing haloes will find their orbital radii
decreasing over time as the depth of the potential well increases.
Figure~\ref{radius_shuffle} shows that this effect (the compaction of the GC
system as the halo grows) is detectable in the change in GC radii, with more GCs
migrating onto smaller (rather than larger or identical) radii between when they
form and when we could observe them at $z=0$.  As the ex-situ clusters are
deposited on relatively larger radii than where they formed, the relative
reduction of in-situ clusters at $r>10\kpc$ is compensated through the ex-situ
clusters, which are distributed nearly uniformly through the halo (as is shown
in Figure~\ref{radius_split}).


\subsection{Birth environment and cluster survival}
The trends we have seen in the $z=0$ GC population are a function of both the
relative efficiencies of different GC formation mechanisms and the ability of
proto-GCs to survive to the present time.  We have already seen in
Figure~\ref{pie_charts} that accreted proto-GCs are more likely to survive than
in-situ proto-GCs. Proto-GCs formed in minor mergers are somewhat less likely to
survive than the average proto-GC (they make up $8.4 \pm 0.3$ per cent of formed
proto-GCs, but only $7.2 \pm 0.5$ per cent of the $z=0$ GCs), while those formed
in major mergers are more likely to survive ($10.9 \pm 0.3$ per cent of
proto-GCs form in major mergers, compared to the surviving fraction of $12.5 \pm
0.7$ per cent of $z=0$ GCs).

\begin{figure}
    \includegraphics[width=\columnwidth]{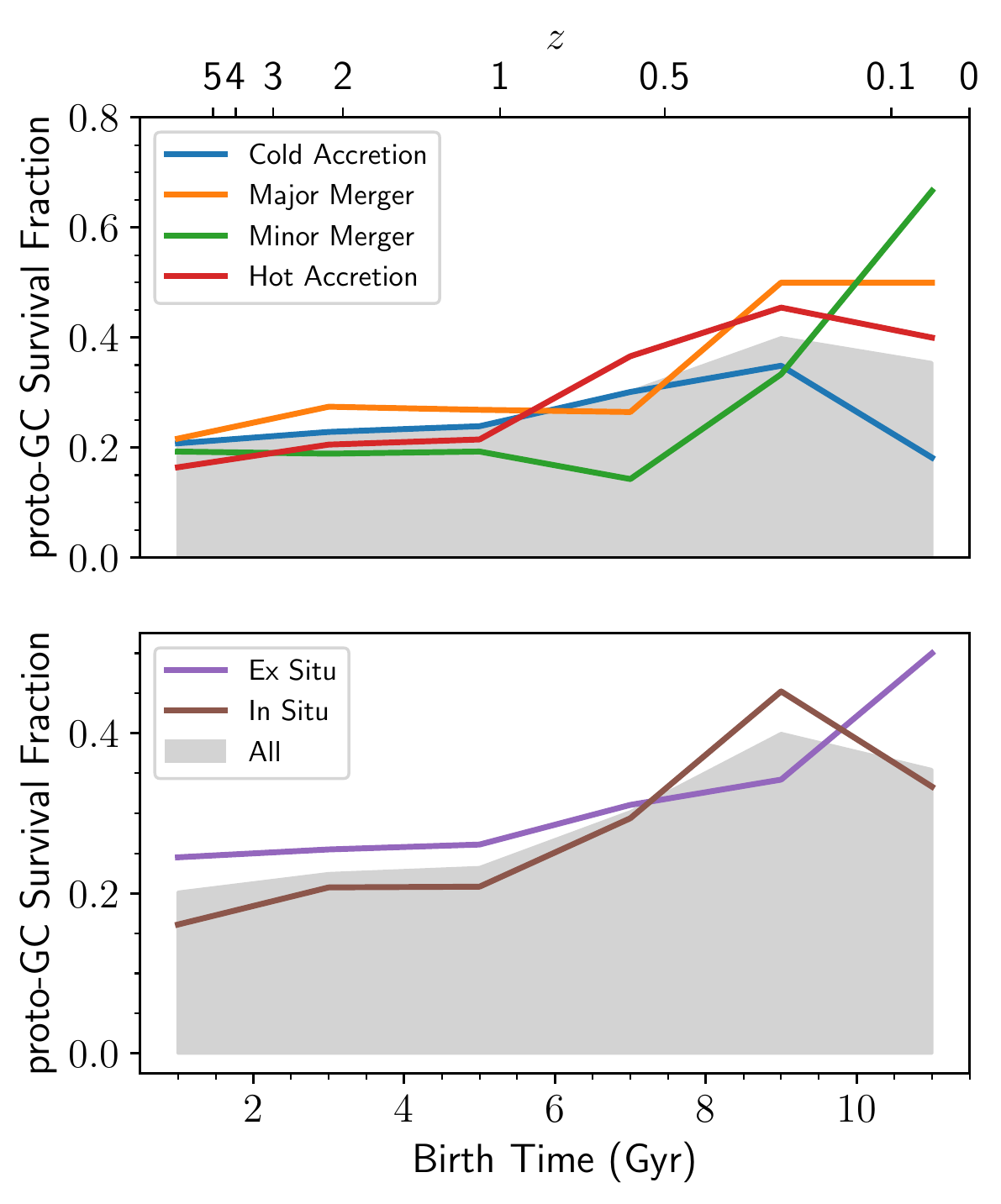}
    \caption{Fraction of proto-GCs that survive to $z=0$ as a function of time.
    For all GC formation mechanisms, we see an increase in the survival fraction
    over time owing to the fact that younger clusters form in a less disruptive
    environment (and have simply had less time to experience mass loss).  As is
    clear from the top panel (survival fraction split based on formation
    mechanism), all formation channels see roughly similar survival rates, with
    major mergers having the highest survival rates during the peak of GC
    formation, $z>1$, and with minor mergers having the lowest survival rate.
    For the same time period, we also see in the bottom panel a higher survival
    rate for ex-situ, accreted proto-GCs.}
    \label{redshift_survival}
\end{figure}

In Figure~\ref{redshift_survival}, we examine the survival rate of proto-GCs
formed in different channels over time.  As we might expect, the average
survival rate increases over time, as younger clusters have simply had less time
to be disrupted than older clusters, and galaxies become less disruptive as they
age and their ISM pressure drops.  For the epoch when proto-GC formation is the
most efficient, $z>1$, we can see that the differences in GC disruption seen in
Figure~\ref{pie_charts} are reflected here. The survival rate of in-situ
clusters is lower than of ex-situ clusters, those born in major mergers have
the highest survival rate, and those born in minor mergers have the lowest
survival rate.  A proto-GC formed during the period of efficient GC formation
had a relatively small chance of surviving to $z=0$, of about $20$ per cent.  This
is reflected in the overall survival fraction of $\sim22$ per cent.

\begin{figure}
    \includegraphics[width=\columnwidth]{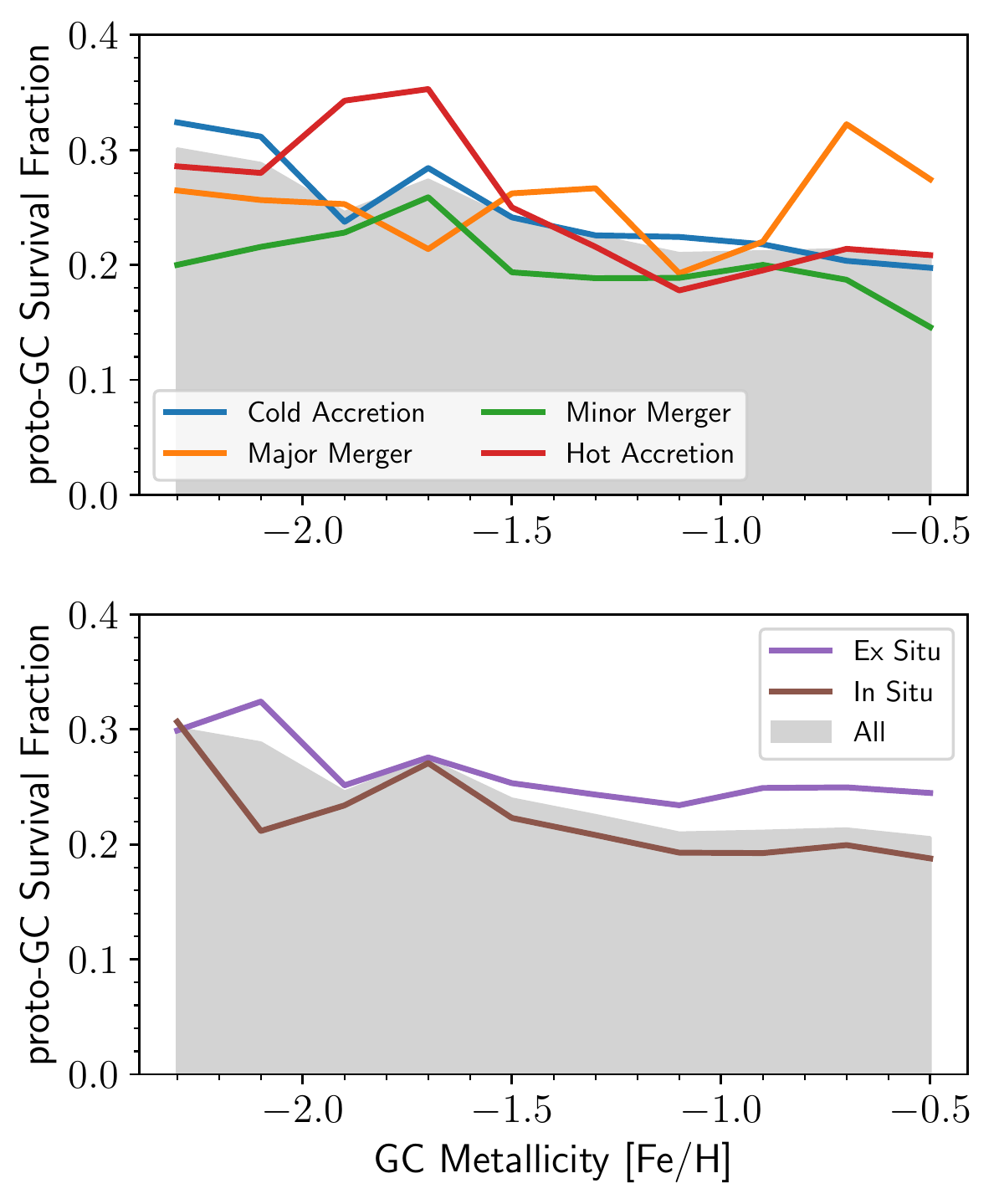}
    \caption{Survival fraction of proto-GCs as a function of their metallicity.
    As with the previous figures, the top panel shows the survival fractions
    split based on formation mechanism, while the bottom panel shows survival
    fractions split based on halo origin (in-situ or ex-situ). For all
    proto-GCs, regardless of their formation mechanism, we see a decrease in the
    survival fraction towards higher metallicities.  Again, we see that
    proto-GCs formed in minor mergers have the lowest survival rates, and those
    formed ex-situ have higher survival rates than those formed in-situ.}
    \label{metallicity_survival}
\end{figure}

The survival fractions as a function of metallicity, shown in
Figure~\ref{metallicity_survival}, also shows the overall higher survival rate
for ex-situ clusters.  We see here as well that metal-rich clusters have lower
survival fractions than metal-poor clusters for all of the populations we
examine, except for a slight bump in the metal-rich clusters formed in major
mergers.  Some of these clusters correspond to the relatively late-forming
population seen in Figure~\ref{redshift_split}, and their high survival fraction
is likely an artefact of their young ages.  When we consider the decreasing
survival rate as a function of metallicity along with the mass-metallicity
relation shown in Figure~\ref{MassMetal}, what we are seeing here is that more
massive haloes (in which more metal-rich clusters form) subject their proto-GCs
to stronger disruption than less massive haloes.\footnote{\citet{Kruijssen2015}
proposed that this trend of disruption rate with metallicity plays an important
role in setting the increase of the GC specific frequency towards low
metallicities and galaxy masses.}

\begin{figure}
    \includegraphics[width=\columnwidth]{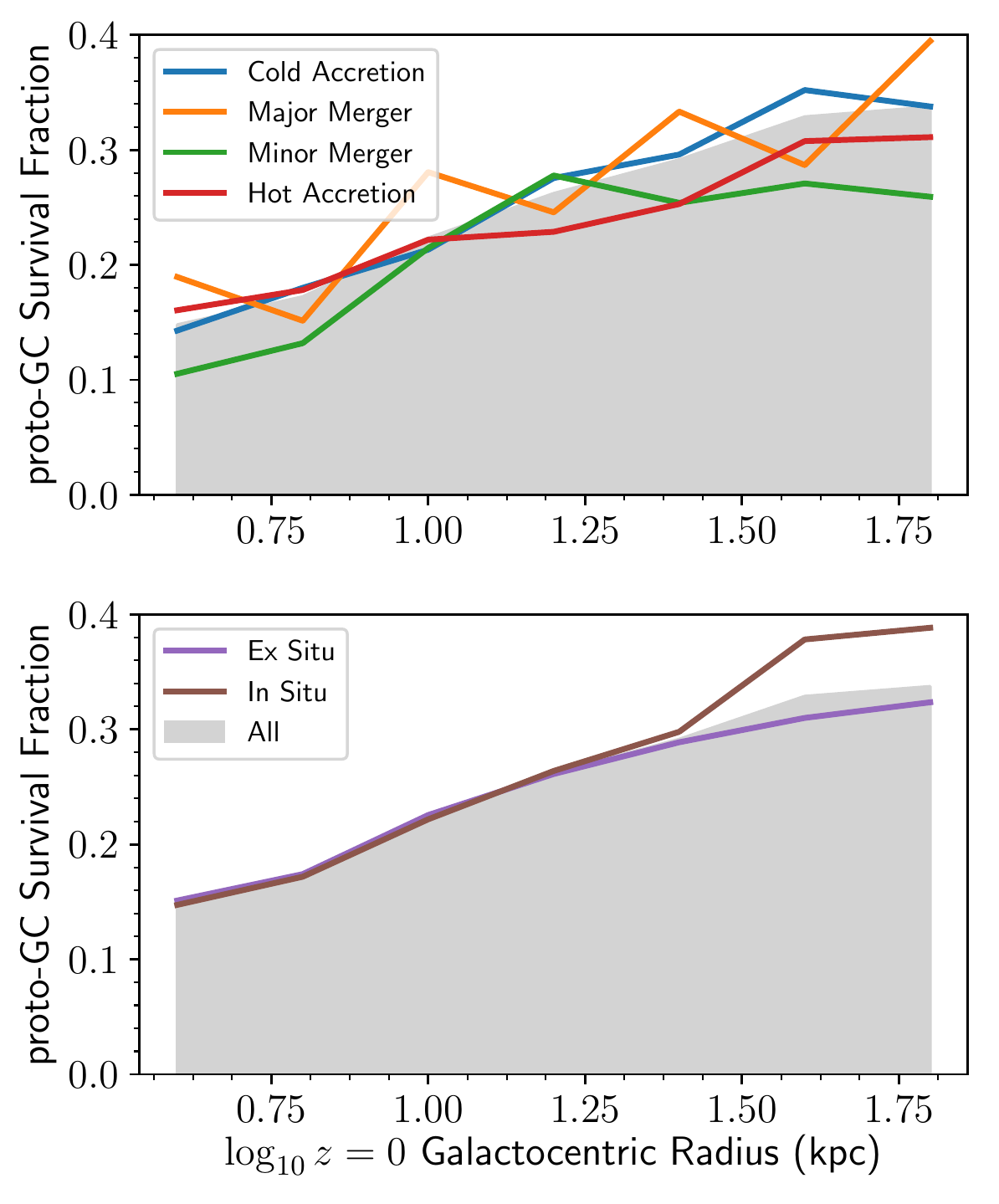}
    \caption{Survival fractions of proto-GCs as a function of their $z=0$
    galactocentric radius.  Proto-GCs that form closer to the centre of their
    parent halo experience stronger tidal interactions, producing a
    significantly rising survival fraction as a function of radius.  Proto-GCs
    that are found at the edge of the halo are $\sim5$ times more likely to
    survive to $z=0$ than those which are found near the galaxy disc.  We see
    little difference in this relation between in-situ and ex-situ clusters,
    while the trend of lower survival rates for proto-GCs formed in minor
    mergers is again evident.  Even for distant halo GCs, the survival fraction
    is $<40$ per cent, a result of the early disruption in the disc, prior to
    their ejection to large galactocentric radii.}
    \label{radius_survival}
\end{figure}

Figure~\ref{radius_survival} shows a significant trend in survival fraction as a
function of the final galactocentric radii at which proto-GCs find themselves.
We use the position of star particles that host disrupted clusters at $z=0$ to
assign the radii of proto-GCs at $z=0$.  Here we can see that we again have a
universal trend in proto-GC survival, regardless of formation mechanism.
Proto-GCs that are found in the stellar halo at $z=0$ have significantly higher
survival rates (roughly a factor of 5 larger) than those found near the galactic
disc.  As proto-GCs are disrupted through tidal interactions (both smooth tides
and tidal shocks), the closer they are to the denser environment of the galactic
disc, the stronger their tidal mass loss will be.  Those GCs we see in the halo
at $z=0$ have likely spent a significant amount of time at large galactocentric
radii, and thus have spent much of their lifetimes in a much gentler tidal
environment than those we see closer to the disc.  The most significant period
of tidal disruption occurs early in the proto-GCs lifetime, when it is still
embedded in the dense ISM.  GCs that spend longer in the disruptive environment
of the ISM prior to ejection will find themselves, on average, at smaller
galactocentric radii (as the potential well will have grown deeper).  

\begin{figure}
    \includegraphics[width=\columnwidth]{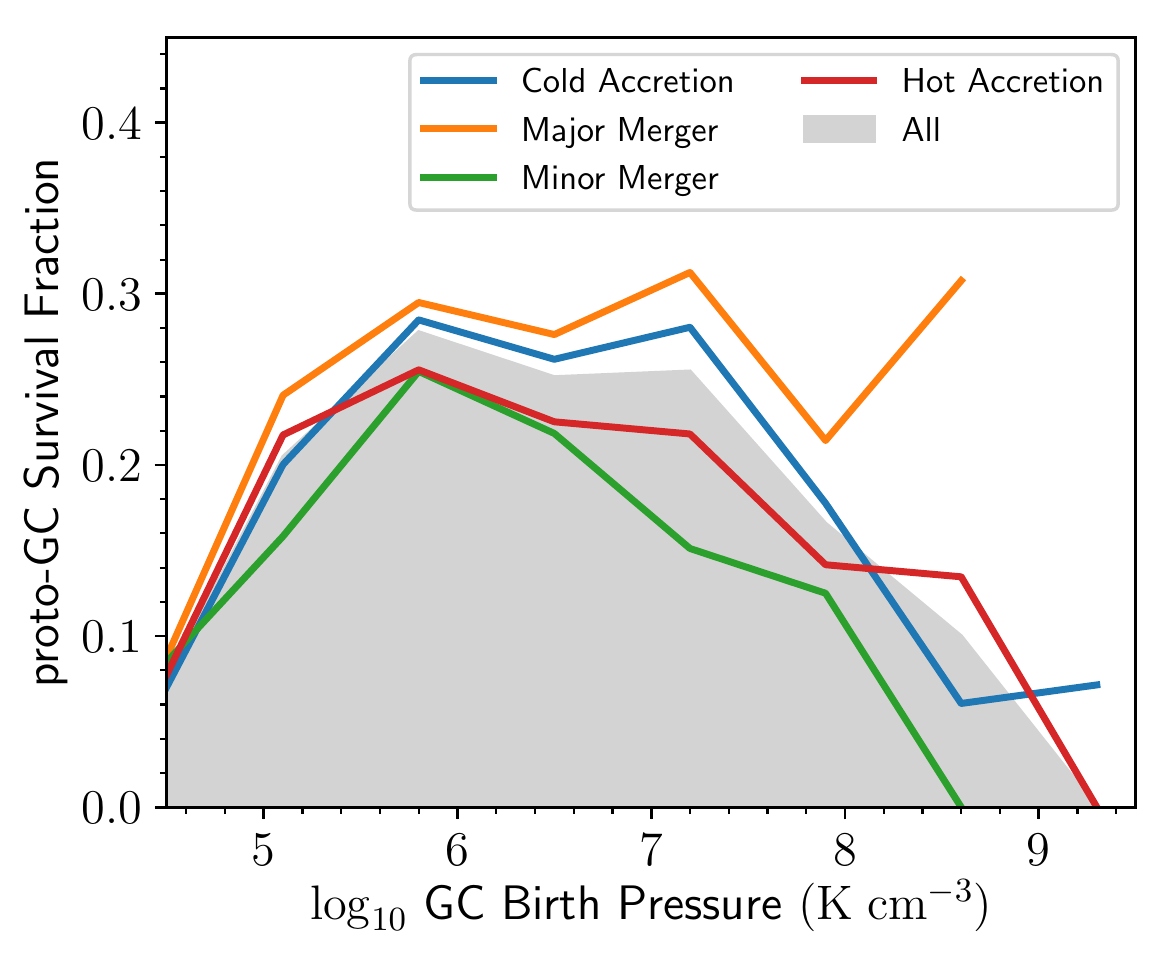}
    \caption{Survival fractions of proto-GCs as a function of birth pressure.
    The decreasing survival fraction, from $\sim30$ per cent at $P/k_{\rm B}\sim10^6\K\hcc$ to
    $\sim5$ per cent at $P/k_{\rm B}\sim10^8\K\hcc$ is evidence of the ``cruel cradle''
    effect.  This process is stronger at higher ISM density and pressure, and as
    a result proto-GCs formed from high pressures are less likely to survive to
    $z=0$, as they are destroyed rapidly by tidal shocks from dense clouds in
    the ISM.  A major reason for the ``bump'' at high pressures seen in the
    proto-GCs formed through major mergers is the formation of more
    GCs from high-pressure gas in low redshift major mergers vs. high-redshift
    major mergers.  As figure~\ref{redshift_survival} shows, younger proto-GCs
    have larger survival fractions, meaning these proto-GCs have a higher
    survival rate simply due to their young age.}
    \label{pressure_survival}
\end{figure}

The disruption of clusters has been predicted
\citep{Gieles2006,Elmegreen2010a,Elmegreen2010b,Kruijssen2011,Kruijssen2015} to
be dominated by tidal shocks in the natal environment, as clusters move through
the dense, high-pressure ISM in which they form.  The strength of these tidal
shocks increases at higher ISM densities and pressures. As
Figure~\ref{pressure_survival} shows, we see exactly this effect in the survival
fraction of proto-GCs as a function of birth pressure.  At higher pressures,
significantly fewer proto-GCs are able to survive to $z=0$, with essentially all
clusters formed at $P/k_{\rm B}>10^9\K\hcc$ being destroyed by tidal shocks.
Interestingly, the only sub-population of GCs that show high survival fractions
when formed from high pressures are those formed during major mergers.  This has
been predicted as a natural outcome of cluster migration
\citep{Kruijssen2011,Kruijssen2012b}.  Major mergers can ``save'' proto-GCs from
disruption by depositing them into tidal tails, transporting them out of the
dense ISM into the halo on a timescale shorter than the timescale required for
disruption by tidal shocks, i.e. $\sim100\Myr$.  This is why we see a larger
overall survival rate for clusters formed in major mergers, especially those
formed at higher ISM pressure.

These trends all help explain the different survival rates we see for each of
the different formation mechanisms in Figure~\ref{pie_charts}.  Ex-situ clusters
have greater survival fractions than in-situ clusters, because they form in
lower-mass haloes (as we would expect from Figure~\ref{metallicity_split} based
on their lower metallicity and the mass-metallicity relation from
\citealt{Erb2006}) and orbit on much larger galactocentric radii (as we see in
Figure~\ref{radius_split}).  Despite forming in lower-mass haloes, proto-GCs
formed through minor mergers are relatively old, and the merger events that form
them were less likely to fling them out to large galactocentric radii than those
proto-GCs formed in major mergers.  This, combined with the population of
relatively young GCs formed in late major mergers gives us the somewhat
surprising result that proto-GCs formed during minor mergers are less likely to
survive to $z=0$ than those formed in major mergers.  The simple trends we see
in age or metallicity for different formation mechanisms are not enough, a
priori, to predict the survival rates of different mechanisms.  The effect of
disruption on proto-GC survival has a non-trivial effect on the distribution of
clusters we see at $z=0$.  The decreasing survival rate as a function of
metallicity has the effect of flattening the metallicity distribution, as the
survival rate as a function of metallicity has the opposite slope to the initial
metallicity distribution.  The same is true for the survival rate as a function
of radius, with the increasing survival at high radii acting to flatten the
radial profile by increasing the disruption rate of the disc clusters.


\section{Do globular clusters form in substructure or minihalo collisions?}
\label{s:collisions}
\begin{figure}
    \includegraphics[width=\columnwidth]{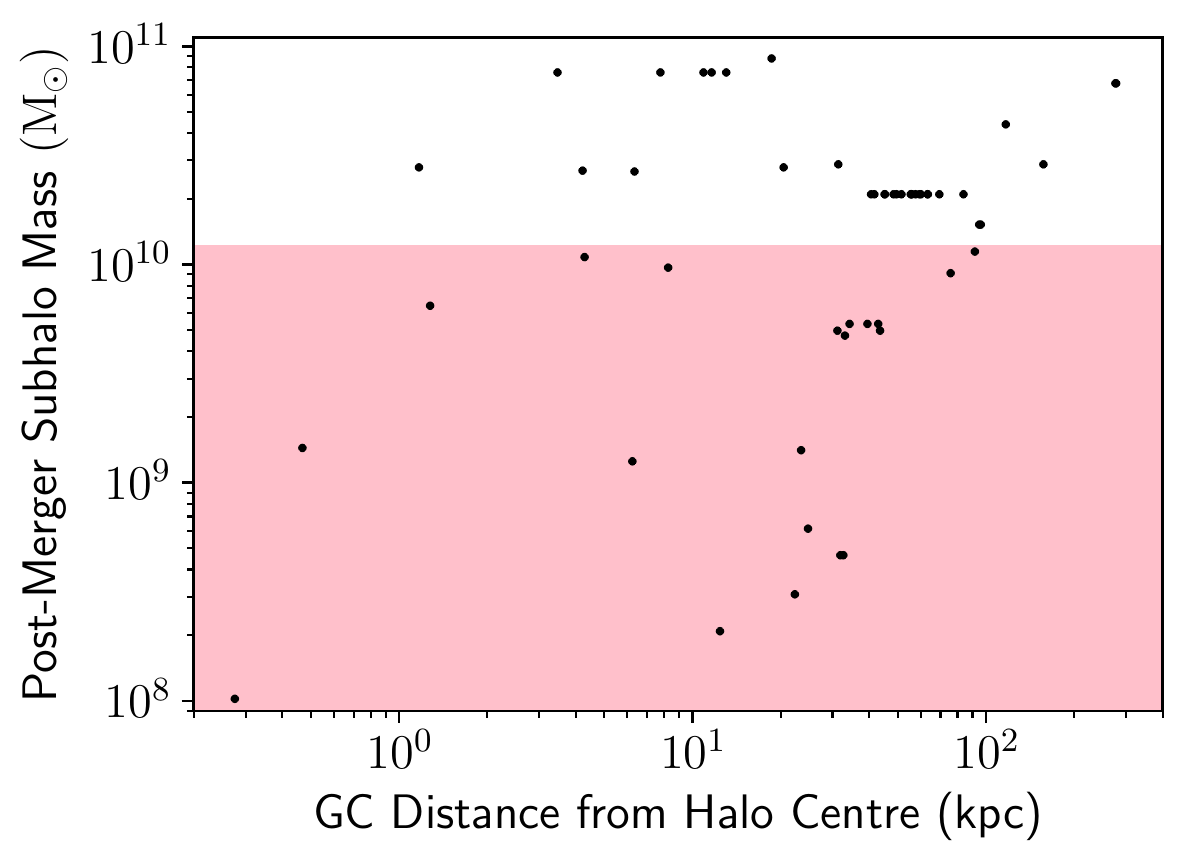}
    \caption{The mass of subhaloes produced by collisions of other subhaloes and
    form GCs as a function of the GC's distance from the centre of potential of
    the primary halo the subhalo resides in.  Most collisions occur between
    $10\kpc$ and $100\kpc$, and the median mass of the post-collision subhalo is
    $2.2\times10^{10}\Msun$.  Haloes affected by resolution (i.e.\ those with
    fewer than $10^4$ dark matter particles) are shown
    in the transparent red region.  A significant fraction of these identified
    collisions are poorly resolved, and may be subject to significant numerical
    error.}
    \label{collision_mass_radius}
\end{figure}
\begin{figure}
    \includegraphics[width=\columnwidth]{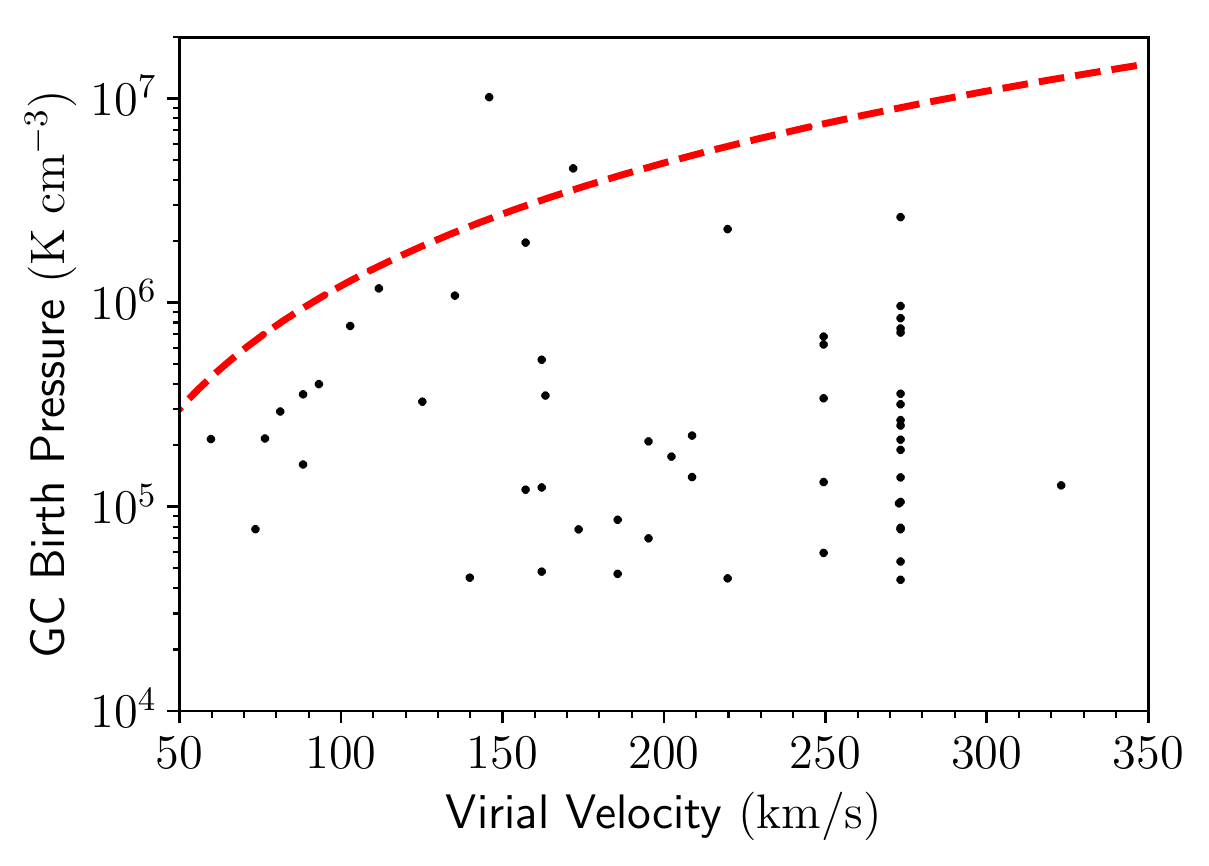}
    \caption{Birth pressure of GCs formed by subhalo collisions as a function of
    the virial velocity of the primary halo that contain those subhaloes.  Black
    points show the true GC birth pressures, while the red dashed line shows the
    ram pressure produced by head-on collisions of subhaloes with ISM density of
    $1\hcc$ at the virial velocity of their parent halo.  Only two clusters are
    formed at higher pressure than can be produced by the collision, whereas the
    majority of clusters can easily form through collisions at lower speed or
    with lower averaged ISM pressure.}
    \label{collision_pressure_velocity}
\end{figure}
One proposed mechanism \citep{Smith1999,Madau2019} for the formation of GCs is
through the collisions between subhaloes which then causes their gas to decouple
from their dark matter (a small-scale analogue of the archetypical Bullet
Cluster scenario, see \citealt{Clowe2004}). These clusters would actually form
far from the disc, without the need for mergers and tidal encounters to deposit
them into the halo and that way ensure their long-term survival.

The recent theoretical work of \citet{Madau2019} has sought to explain the
specific observational facts we now have for the GC systems in the MW and
M31.  Namely, that observations of GCs in the outskirts of the galaxy halo, such
as MGC1 have put strong constraints on the amount of dark matter
these clusters may contain, consistent with them lacking any dark halo
at all \citep{Conroy2011}.  \citet{Madau2019} propose that a simple model
for producing these dark matter free halo GCs that relies on two basic
assumptions: that GCs form when gas pressures reach $P/k_{\rm B} \sim 10^6-10^7 \K\hcc$,
and that these pressures can be achieved through the collision of cool atomic
clouds with densities of $\sim\hcc$ in substructure of the haloes of $L^*$
galaxies, where orbital velocities of $\sim 200\kms$ are sufficient to produce
these conditions through ram pressure.  The high pressures assumed for this
model are justified by the fact that above $10^7\K\hcc$, the CFE approaches
unity. The authors estimate, based on analytic kinetic theory and N-body simulations,
that a few hundred of these collisions will occur within the halo of $L^*$
galaxies, most of which will have relative velocities sufficient to produce the
high pressures they require. Of course, the alternative hypothesis for
explaining the above observations is the one we obtain from E-MOSAICS, namely
that GCs form as the natural outcome of high-redshift star formation in a way
that is fundamentally unassociated with dark matter.

A similar proposed mechanism, albeit at higher redshift, is the collision of
atomic cooling minihaloes (with virial masses $\sim10^8\Msun$) at high redshift,
prior to the formation of stars within said haloes \citep{Trenti2015}.  In this
mechanism, GCs form as the nuclei of these minihaloes after a major merger, with
a dark matter halo surrounding them, and with re-enrichment of a second
generation of stars fed via slow stellar winds.  The dark matter halo is
subsequently stripped away by the assembly of the main halo they fall into.
While this chemical enrichment scenario is not one we can probe with the single
phase of cluster formation intrinsic to the MOSAICS model, we can quantify the
frequency of GC formation in minihalo collisions.

Since the mechanisms proposed by \citet{Trenti2015} and \citet{Madau2019} rely
only on simple hierarchical structure formation and hydrodynamics, it should in
principle manifest itself in the E-MOSAICS simulations, with some important
caveats. Firstly, the E-MOSAICS simulations are run at a significantly lower
dark matter resolution than the {\it Via Lactea I} simulations studied in
\citet{Madau2019} ($1.6\times10^6\Msun$ as opposed to $2.1\times10^4\Msun$
respectively) or the simulations used in \citet{Trenti2015} (which used a mass
resolution of $8.3\times10^4\Msun$).  This means that many of the subhaloes we
identify, as well as the majority of the smaller pairs identified as colliding
in \citet{Madau2019} are below the resolution limit of $10^4$ dark matter
particles.  These subhaloes will have dynamics significantly affected by both
softening and discreteness noise.  Secondly, while the \citet{Madau2019}
analysis uses the high time resolution ($68.5\Myr$ between snapshots) to
identify close passages of subhaloes, we lack a sufficiently high cadence of
simulation outputs to follow this method.  Instead, we rely on the merger tree
generated using subhaloes of bound particles identified by {\sc SUBFIND}.  This
means we may potentially miss events that occur between snapshots, if these
smaller subhaloes are able to both accrete and collide between outputs.  It also
may mean these results are sensitive to the details of subhalo identification.
While {\sc SUBFIND} has been shown to produce good results for subhalo
identification that include baryons \citep{Dolag2009}, the fact that many of our
identified collisions involve subhaloes near the resolution limit may limit the
accuracy of the algorithm.

We can look at our population of GCs to determine how many (if any) are formed
through the substructure collision mechanism.  As we saw in
Figure~\ref{pressure_split}, most clusters form at pressures much lower than the
$10^7 \K\hcc$ that \citet{Madau2019} invoke substructure collisions to produce.
There is, however, a tail of clusters produced from ISM pressures $>10^7
\K\hcc$, and substructure collisions still may produce GCs from lower-pressure
gas if they have lower initial density or relative velocities.  We find that out
of the 2732 GCs in E-MOSAICS, 613 form in substructure: {\sc SUBFIND} subhaloes
that are within the virial radius of larger halo.  From these, we can select
substructure that is formed from the merging of two previous subhaloes
containing bound gas, and find 54 candidate GCs formed through the merging of
substructure across our full sample of galaxies ( $\sim2$ per cent of the total
GC population, or about 1-3 GCs per halo).  Note that here we do not restrict
our sample to mergers that result in dark matter-deficient objects (where the
collisionless component has decoupled from the gas), so the numbers we see here
can be considered an upper limit for the number of objects formed through this
route. As \citet{Trenti2015} require minihalo collisions with mass ratio
identical to what we have used for our major merger criterion (mass ratio
$>0.25$), we can begin with the 343 GCs that have formed in major mergers.  Of
these, only 27 form from mergers between star-free halos.  Of this small
fraction, only a single one forms prior to $z\sim6$, when atomic cooling can
proceed without the heating from the post-reionization UV background.

In Figure~\ref{collision_mass_radius}, we examine where these subhalo collisions
occur and see that the majority of the GCs formed in post-merger subhaloes are
found between $10\kpc$ and $100\kpc$ from the main halo's centre of potential.
Many of these subhaloes are around the peak masses identified in
\citet{Madau2019} of colliding {\it Via Lactea} simulated subhaloes
($\sim2\times10^{10}\Msun$), but a significant number are also fairly massive,
approaching $10^{11}\Msun$.  These are likely clusters formed in the tidal tails
and debris of major mergers, when the primary halo contains much of the
substructure of larger, accreted haloes.  We can also look directly at the
proposed mechanism of the \citet{Madau2019} GC formation scenario, at the birth
pressure of GCs formed in subhalo mergers as a function of the virial velocities
of the primary halo they live in.  In Figure~\ref{collision_pressure_velocity},
we show the birth pressures for all 54 GCs formed in substructure collisions, as
well as the maximum ram pressure produced by two subhaloes colliding head-on,
each at the virial velocity of the primary halo.  All of the GCs (except for 2)
formed through substructure collisions form with birth pressures below this
maximum ram pressure for gas at a density of $1\hcc$ ($P_{\rm ram}=\rho
V_{200}^2$, where $V_{200}=\sqrt{2GM_{200}/R_{200}}$).  Thus, it is conceivable
that the collisions of these subhaloes, with lower relative velocities or
densities, or with some amount of radiative cooling prior to GC formation can be
explained solely through the ram pressure of the collision.  The E-MOSAICS
simulations thus do contain GCs formed through the mechanisms proposed by
\citet{Trenti2015} and \citet{Madau2019}, albeit a very small fraction ($\sim2$
per cent for subhalo collisions, and $\sim1$ per cent for minihalo major
mergers) of the total GC population.

\section{Discussion}
\label{s:discussion}
\subsection{The conditions of GC formation}
The results presented in this paper show that the picture of GC formation is
many-faceted.  Single mechanisms, such as gas-rich major mergers, may account
for a non-trivial fraction of GCs formed, but the conditions required for GC
formation are not so difficult to achieve as to exclude other mechanisms, from
the collision of subhaloes to the simple collapse of massive GMCs in gas-rich,
turbulent discs at high redshift.  What establishes the fractions of GCs formed
through each of the channels examined here is the relative frequency of gas
elements reaching the high pressures and densities required for clusters to
form, combined with the subsequent evolution of that cluster allowing it to
survive to $z=0$.  The typical ISM pressures from which GCs are formed, as we
showed in Figure~\ref{pressure_split} are relatively high, with $P/k_{\rm B}
\sim10^{5.5}\K\hcc$, but not quite as extreme as some proposed requirements
($P/k_{\rm B}>10^7\K\hcc$ in \citealt{Madau2019} for example).  This means that
the ISM of high-redshift galaxies can frequently reach these pressures without
extreme events or exotic physics \citep[see][]{Elmegreen2010b,Kruijssen2015}.
Despite the relatively low cluster formation efficiency at these pressures
($\sim10$ per cent, \citealt{Pfeffer2018}), the much higher frequency with which
regions of the ISM can reach these pressures means that only $8.0 \pm 1.3$ per cent of
the surviving GCs formed with $P/k_{\rm B}>10^7\K\hcc$.  The cold, clumpy,
turbulent environment we identify as the primary site of GC formation may even
produce GCs within the filaments that feed galaxies, prior to the actual
accretion of this material \citep{Mandelker2018}.  While we lack the resolution
in E-MOSAICS to resolve the fragmentation of accretion filaments, future work at
higher resolution may be able to disentangle whether the clusters we identify as
forming through cold form within the turbulent disc, or within fragmentation of
cold filaments within the halo.

Dynamical disruption is clearly an important factor in shaping the GC
populations we see at $z=0$, as $\sim80$ per cent of proto-GCs (defined as
having an initial mass larger than $10^5\Msun$) formed in E-MOSAICS do not
survive to $z=0$.  Despite the importance of dynamical disruption, the formation
channel of a proto-GC seems to have little impact on whether it will survive to
the present.  Ex-situ clusters do experience less disruption compared to in-situ
ones, and we find that low metallicity clusters, along with those that end up at
larger ($z=0$) galactocentric radii have higher survival rates compared to those
with high metallicity or low galactocentric radii.  Notably, we also find that
proto-GCs formed from extremely high pressures $P/k_{\rm B}>10^8\K\hcc$ are
almost universally destroyed, except for those that are formed in major mergers,
because they are efficiently ejected from their disruptive birth environments
\citep[also see][]{Kravtsov2005,Kruijssen2012b}.

All of this suggests that the E-MOSAICS GC population is well-described by the
picture of GC formation presented in \citet{Kruijssen2015}:  proto-GCs are
efficiently formed, and destroyed, in the high-pressure, gas-rich discs of high
redshift galaxies.  Those that we see today are those that have survived by
being either ejected from the in-situ disc, or delivered from an ex-situ disc,
during subsequent hierarchical merging.  The high-redshift galaxies in which GCs
form have high gas fractions, as we see in Figure~\ref{GasFrac}, and has been
observed e.g.\ \citet{Tacconi2013}.  Fueled by cold filaments, these clumpy,
turbulent, gas-rich discs are sites of efficient proto-GC formation.  The
conditions that make these galaxies ideal to form clusters also make them
efficient destroyers of clusters, stripping the mass from proto-GCs through
tidal shocks as they move through this dense ISM.  However, the high frequency
of mergers at high redshift can act to expel proto-GCs onto high radii, where
they will evolve more slowly under the influence of weaker tidal evaporation.
This is clearly revealed through the survival rates for clusters formed at
different pressures in Figure~\ref{pressure_survival}.  Only when a major merger
occurs a short time after formation, or when a cluster forms during this merger,
can a proto-GC formed in a $P/k_{\rm B}>10^8\K\hcc$ ISM be ejected from that ISM
quickly enough.  The importance of tidal shocks from the ISM is discussed in
more detail in the analysis presented in Appendix~\ref{LiComparison}, where we
compare the disruption mechanisms of E-MOSAICS and a recent, similar set of
cosmological simulations \citep{Li2017a,Li2018,Li2019a}.  The analysis presented
in Appendix~\ref{LiComparison} shows that a distinguishing feature of proto-GCs
that survive to $z=0$ is that they have experienced very little mass loss from
tidal shocks.  Essentially all $z=0$ GCs have lost less than $10^5\Msun$ due to
tidal shocks, despite the fact that most proto-GCs experience at least this much
shock-driven mass loss.

\subsection{Comparison to observations}
A great deal of comparisons have been made between the E-MOSAICS simulations and
observations of the MW and other local galaxies.  \citet{Pfeffer2018} showed
that the E-MOSAICS galaxies, in general, match the specific star formation
rate of the MW back to $z=6$, as well as the high-mass end ($>10^5\Msun$) GC
mass function and maximum mass to galactocentric-radius relation in the MW at
$z=0$.  A more comprehensive set of observational comparisons were made in
\citet{Kruijssen2019a}.  Here, it is shown that the metallicity distribution,
spatial density profile, and specific frequency-stellar mass relationship
match observations of the MW, M31, and Virgo Cluster galaxies.  In particular,
the metallicity and radial distribution we show in
Figure~\ref{metallicity_split} and Figure~\ref{radius_split} were previously
examined in \citet{Kruijssen2019a}.  While that study did not look at the split
based on formation mechanism as we do here, they did find that the metallicity
distribution in the E-MOSAICS galaxies is consistent with the MW
\citep{Harris1996} and M31 \citep{Caldwell2011}.  The radial profiles of GCs is
also compared with the MW density profile fit by \citet{Djorgovski1994}.

As many studies \citep{Blakeslee1997,Spitler2009,Burkert2020} have found, there
is a tight relation between the number of GCs and halo mass across 6 dex in halo
mass.  While the E-MOSAICS simulated galaxy sample is currently limited to 25
MW-mass $L^*$, objects (these are the locations of most $z=0$ GCs, as shown by
\citealt{Harris2016}), we have also produced a simulated $34\;\rm{cMpc}$
cosmological volume.  Analysis of this volume has allowed us to probe the
$N_{\rm GC}-M_{\rm halo}$ relation across a much wider dynamical range
\citep{Bastian2020}.  

A wide variety of observational constraints from the Local Group deal with the
internal evolution and properties of GCs.  Whether this comes in the form of
evidence for multiple populations \citep{Bedin2004,Renzini2015,Bastian2018}, internal
kinematics \citep{Watkins2015,Kamann2018,Bastian2018,Baumgardt2019}, or mass segregation
\citep{Baumgardt2008,Beccari2010,Webb2017}, all of this occurs at physical
scales well below our resolution.  However, many of these constraints can be
used as indirect probes of the formation mechanisms and history of GCs.  For
example, a recent analysis of GC phase-space data from Gaia data by
\citet{Baumgardt2019} have suggested that the MW may have formed $\sim 500$
proto-GCs, consistent with the average of 487 proto-GCs formed per E-MOSAICS
galaxy.  Observations of the mass function of GC stars \citep{Sollima2017} have
been used to infer that MW GCs have lost $\sim3/4$ of their mass since
formation \citep{Webb2015,Kruijssen2015,Baumgardt2017}, consistent with the mass loss found in
E-MOSAICS by \citet{Pfeffer2018} and \citet{Reina-Campos2018}.

One limitation of current observational constaints for GC properties is that
they (nearly all) come from nearby, low-redshift systems.  As a result, we do
not yet know how the $N_{\rm GC}-M_{\rm halo}$ relation evolves with time to act
as a comparison to the predictions from E-MOSAICS
\citep{Bastian2020,Kruijssen2020}.  However, for GCs in the Local Group, stellar
age dating can give us some idea as to the overall formation history of the GC
systems in this handful of galaxies.  For younger, more metal-rich GCs,
age-dating can provide relatively tight constraints on the formation redshift
(for example, the SMC GCs NGC 339, NGC 416, and Kron 3 all formed at
$z\sim0.65$, \citealt{Niederhofer2017}).  At lower metallicity and greater age,
uncertainties in age estimates become more significant.  The age estimates for
47 Tuc determined in \citet{Hansen2013}, calculated using the cooling of its
white dwarf populations, yield an age of $9.9\pm0.7\Gyr$.  Studies of the GC
ages in the MW \citep{VandenBerg2013,Leaman2013} and M31 \citep{Caldwell2011}
have found the bulk of GC ages between $10-13\Gyr$, with individual cluster age
uncertainties of $\sim0.5\Gyr$, broadly consistent with the ages we find here
and in \citet{Reina-Campos2018} and \citet{Kruijssen2019c}.  Other methods for
age dating give comparable uncertainties for old stellar populations
$(\sim1\Gyr)$ (a detailed review of this can be found in Section 5.4 of
\citealt{Kruijssen2019c}.  This means that a cluster formed at $z\sim6$ is
difficult to distinguish observationally from one formed at $z\sim3-4$.
Compounding this uncertainty is the requirement to resolve individual stars for
most accurate age-dating techniques.  This limits the sample of galaxies with
accurate GC population ages to a the Local Group, which may bias our
observational picture of when the ``typical'' GC forms.  Observations of nearby
early-type galaxies suggest that they may contain a younger GC population
compared to the MW \citep{Usher2019}.

The problem of identifying high-redshift proto-GCs has been approached along a
number of angles.  \citet{Renzini2017}, \citet{Boylan-Kolchin2018}, and
\citet{Pozzetti2019} have all identified that the brief, intense star formation
that forms GCs should be detectable by the James Webb Space Telescope (JWST),
even up to $z=10$.  The number of detectable young proto-GCs will provide
constraints on the initial masses of proto-GCs and their cosmic formation
history.  As we have shown here, and as was previously discussed in detail in
\citet{Reina-Campos2019}, the E-MOSAICS model predicts that most GCs form
between $z=2-4$, well within the capabilities that \citet{Renzini2017},
\citet{Boylan-Kolchin2018}, and \citet{Pozzetti2019} predict for JWST.  This
would allow us to directly compare models that form most GCs early, at $z>4$,
to models with more extended epochs of GC formation, like E-MOSAICS.  These
measurements would be independent of the uncertainties involved in age-dating
old GCs, as the UV luminosity of stellar populations falls precipitously after
a few tens of Myr.  \citep{Pfeffer2019a} has examined the UV luminosity
properties of high-redshift proto-GCs in the E-MOSAICS simulations, and future
observations will allow us to test these predictions.

\subsection{Comparison to other simulations}
E-MOSAICS is not the first simulation suite, cosmological or otherwise, to
investigate the formation of GCs.  Two of the earliest hydrodynamic simulations,
\citet{Bekki2002a} and \citet{Bekki2002b}, used SPH simulations  of mergers
between dwarf \citep{Bekki2002a} and $L^*$ \citep{Bekki2002b} galaxies.
\citet{Bekki2002b} found that low redshift major mergers would produce GC
systems with super-solar metallicity, in disagreement with the low observed
median metallicities of GC systems in $L^*$ and larger galaxies, but that
metal-poor clusters do end up on larger radii than metal-rich ones.
\citet{Bekki2002a} found that GCs produced in minor mergers between dwarf
galaxies are sensitive to the details of the merger, including the orbital
configuration and mass ratio.  These idealised mergers used a simple model for
the formation of GCs, where star forming gas has a fixed, $10$ per cent
probability of forming a GC when the ISM pressure exceeds $2\times10^5 \K\hcc$.
These simulations omit any form of cluster disruption and use a very simple
models for the ISM, with a $10^4\K$ isothermal equation of state, and a gas mass
resolution of $3\times10^6\Msun$ (compared to $2.25\times10^5\Msun$ in
E-MOSAICS).  A similar set of simulations by \citet{Li2004} was performed at 100
times higher resolution, with a different criterion for GC formation (gas
density exceeding $1000\hcc$), and using accreting sink particles to model GCs
embedded in dense molecular gas.  They find that mergers can increase the GC
formation rate by a factor of $\sim3$ over $5\Gyr$, but without any disruption
mechanism it is difficult to say what fraction of these proto-GCs would survive
for a significant time post-merger.  Recently, the approach of simulating
isolated dwarf galaxies has been pushed to mass resolutions of $4\Msun$ by
\citet{Lahen2020}, which allow them to study the formation of individual,
resolved massive stars, and look at the formation of clusters from first
principles in a galactic environment.

The earliest attempt to simulate the formation and evolution of GCs in a
cosmological environment came through \citet{Kravtsov2005}.  These adaptive mesh
refinement (AMR) simulations include a number of improvements over previous
works, beyond the inclusion of a full cosmological history.  These simulations
include feedback and metal enrichment from supernovae, as well as tabulated
density-dependent gas heating and cooling.  However, these simulations do not
follow the evolution of the $L^*$ galaxy to $z=0$, but focus on the early,
$z>3$, evolution.  To identify the sites of cluster formation, the authors built
a cloud catalog of GMCs with densities exceeding $40\hcc$, and pressures of
$>10^4k_{\rm B} \K\hcc$.  Within these clouds, a number of simple analytic
assumptions were made to estimate the mass and radius of clusters formed in the
densest cores.  \citet{Kravtsov2005} find a qualitatively similar distribution
of GC galactocentric radii and metallicities to the analysis presented here, but
given the diverse GC systems seen in E-MOSAICS, it is difficult to interpret the
differences seen in a single object at $z>3$ to the many GC systems we have
simulated to $z=0$.  As the simulations by \citet{Kravtsov2005} were focused on
formation of GCs, they do not include mechanisms for cluster disruption and
evolution.  

More recent attempts to simulate the formation of GCs at high mass resolution of
$\sim10^2\Msun$ have been attempted by \citet{Kimm2016}, \citet{Kim2018b}, and
\citet{Ma2020}.  \citet{Kimm2016} studied the evolution of atomic-cooling halos
to $z=10.2$.  As expected, many of the stars within these GCs are quite metal
poor, with a large fraction of the stars having metallicity $Z/Z_\odot<-4$,  and
with a spread in metallicity of over 4 dex.  These clusters do show a relatively
uniform age for their stellar populations, with most clusters expelling all star
forming gas by SN feedback within $\sim 10\Myr$.  The high resolution that these
simulations used only allowed them to study a pair of GCs, and only for a very
short, early slice of cosmic time.  Work by the FIRE collaboration, recently
reported by \citet{Kim2018b} and \citet{Ma2020}, has also found GC candidates in
high-resolution resimulations of $L^*$ progenitors run to $z=5$.  These
cosmological simulations allowed \citet{Kim2018b} to study the formation, early
rapid mass loss, and longer ($\sim300\Myr$) evolution of a bound cluster in a
realistic progenitor galaxy.  They found that tidal shocks can be a powerful
source of mass loss, as well as a filtering process that removes the least bound
outer stars of the cluster. \citet{Ma2020} looked at the formation sites of
these clusters, finding that bound clusters form preferentially at higher
density (and therefore pressure) than unbound associations or isolated stars,
consistent with the cluster formation of \citet{Kruijssen2012b} that is adopted
in E-MOSAICS.  They also identified that their bound cluster mass function
follows a $-2$ power-law slope, consistent with both observations and the
cluster mass function in E-MOSAICS \citep{Pfeffer2018}.  They also identified
that, in high-resolution simulations, the details of the star formation model
can have a large impact on the density at which stars are formed (similar
results have been identified by \citealt{Kay2002}, \citealt{Hopkins2012a}, and
\citealt{Gensior2020}, among others).  

These types of high-resolution, short timescale ($\ll t_{\rm Hubble}$)
simulations are an important complement to results from the E-MOSAICS simulation
we have shown here.  Much of what such high-resolution simulations are
able to examine (the internal structure of proto-GCs, their detailed formation
process, and the evolution of individual stars) cannot be probed by E-MOSAICS,
as these processes all occur in parameterised sub-grid models below our
resolution scale.  On the other hand, these simulations look at only a handful
of objects: two proto-GCs in the case of \citet{Kimm2016}, a single cluster
examined in detail by \citet{Kim2018b}, and a few hundred clusters in four
objects, evolved for only a few hundred Myr in \citet{Ma2020}.  This prevents
these studies from being able to examine either the long-term evolution of
individual GCs, or the diversity in GC population.  With 25 $L^*$ galaxies,
evolved to $z=0$, E-MOSAICS is designed specifically to look at these important
features of GC evolution.

Only one other set of cosmological simulations takes both cluster formation
physics and the subsequent tidal evolution into account, including both an
observationally-justified mechanism for GC formation as well as self-consistent
disruption. These are the simulations first presented in \citet{Li2017a}.  These
simulations are quite similar to E-MOSAICS, but with a number of critical
differences, and we describe in detail the similarities and differences in
Appendix~\ref{LiComparison}.  The simulations of \citet{Li2017a} have higher
resolution than the E-MOSAICS simulations, but at the cost of being able to
simulate only a single galaxy, evolved only to $z=0.6$.  The higher resolution
of these simulations, combined with the differences in both the hydrodynamic
method and subgrid physics for star formation and feedback make them an
important complementary study to the results that we have presented here.

A number of subsequent simulation studies have taken a post-processing approach
of ``painting on'' GCs to star or dark matter particles with certain formation
criteria.  Whether these criteria are simply based on stellar age
\citep{Halbesma2019}, halo properties \citep{Griffen2010,Ramos-Almendares2019},
or the more realistic ISM conditions used previously, each of these approaches
will suffer from the same critical weakness, illustrated by the difference we
see between proto-GCs (without disruption) and GCs that survive to $z=0$:
roughly $\sim80$ per cent of proto-GCs that form are disrupted before they can
be observed at $z=0$.  As this disruption depends on the precise history and
environment of individual clusters \citep{Reina-Campos2018, Reina-Campos2019},
disruption is a critical piece of the physical picture that produces the $z=0$
GC population.  Because this disruption is highly variable on short timescales
\citep{Pfeffer2018,Li2018}, this effect cannot be simply calculated in
post-processing, without the storage of a prohibitively large number of
snapshots.  This was attempted in a cosmological simulation of a MW-mass galaxy
by
\citet{Renaud2017}.  Their AMR zoom simulations of the FIRE halo ``m12i'' from
\citet{Hopkins2014} has similar resolution to the E-MOSAICS haloes (a minimum
physical cell size of $218\pc$, and includes the comprehensive feedback physics
first presented in \citet{Agertz2013}.  Unlike E-MOSAICS, these simulations do
not include a physical model for cluster formation or disruption, instead opting
to define globular cluster candidates as star particles formed before a lookback
time of $10\Gyr$.  Only a subset of $15,000$ star particles have on-the-fly
tidal tensors calculated, in order to reduce the computational and storage
costs, but these tidal tensors are not used to model any mass loss of the
clusters, and are derived to only measure the contribution of the large-scale
tidal field.  \citet{Renaud2017} find that their simulation reproduces the
metallicity bimodality of the GC population through the difference in
metallicity distribution for in-situ and ex-situ (i.e.\ accreted) clusters
(similar to what we see in Figure~\ref{metallicity_split}).  Their analysis does
show that the tidal fields experienced by GC candidates evolve over time.
However, because they omit the contribution of ISM-driven tidal shocks, and
because they examine only a single object (rather than the 25 we study here),
they are not able to study the diverse evolution of tidally-induced mass loss
we have examined here (see also the detailed analysis of dynamical disruption in
\citealt{Reina-Campos2018}).  

A recent study by \citet{Carlberg2020} presented a nearly opposite approach to
``painting on'' GCs in a cosmological simulation.  Instead, semi-resolved
($5\Msun$ star particle mass) clusters are created with a King profile scaled to
the tidal radius, and placed in a disk-like distribution throughout
\citet{Hernquist1990} halos generated to match the halo catalog of Via Lactea
II, at $z=8$.  These N-body only simulations are then evolved to $z=0$, along
with a Monte Carlo model for internal many-body heating of the cluster.  This
approach allows a more detailed study of the tidal evolution of individual
clusters, since they are better resolved than in E-MOSAICS or the simulations by \citet{Li2017a}
and are evolved to $z=0$, unlike in \citet{Kimm2016} or \citet{Ma2020}.
However, since the clusters themselves are evolving in a dark matter-only
simulation, and are initialised explicitly by being placed in the initial conditions
of the simulation, this approach cannot provide much insight to the formation
mechanisms or history of GC populations.  Like many other studies, it also
provides a 
look at a single $L^*$ galaxy, and thus cannot probe the variety of formation
and assembly histories that a larger sample can.

\section{Conclusions}
\label{s:conclusions}
With E-MOSAICS, we have used cosmological zoom-in simulations of Milky Way-mass
galaxies to examine the formation and evolution of globular clusters in $L^*$
galaxies.  We see some notable trends and differences in the GCs that are
formed in situ or ex situ, as well as those formed through four different
formation channels (hot accretion, cold accretion, major mergers, and minor
mergers).  A summary of the features we have found in the birth environments of
GCs and their survival over cosmic time are as follows.

\begin{itemize}
    \item The GC systems of $L^*$ galaxies are formed through a mixture of
        clusters formed in-situ and those formed ex-situ that are later
        accreted.  Most GCs formed initially in turbulent, high-redshift discs, with
        a small fraction formed during mergers.  This picture of GC formation
        can explain the main features of $L^*$ GC systems without relying on
        physics beyond ``simple'', environmentally dependent star and cluster formation.
    \item While major mergers do produce some ($12.6 \pm 0.6$ per cent) of the GCs in
        $L^*$ galaxies, these GCs are a definite minority of the total
        population.
    \item The vast majority ($77.6 \pm 1.0$ per cent) of proto-GCs are disrupted
        before $z=0$.  This disruption is due to a combination of tidal shocks
        experienced in the ``cruel cradle'' of the birth environment and slower
        evaporation in the halo.  
    \item In-situ clusters are more effectively disrupted than ex-situ clusters,
        but still make up $52.0 \pm 1.0$ per cent of $z=0$ GCs.  Ex-situ clusters slightly
        outnumber in-situ clusters for low metallicities ($\rm [Fe/H] < -1.5$)
        and large galactocentric radii ($r > 10\kpc$).
    \item There is no simple set of criteria to fully isolate, in
        age-metallicity-position space, GCs that formed through any specific
        mechanism, or to determine whether those clusters formed in-situ or
        ex-situ. The reason is that GCs mix relatively well in configuration
        space over cosmic time, due to hierarchical galaxy assembly.
    \item GC metallicity is a good estimate for the stellar mass of the galaxy
        they formed within, which is a simple consequence of the
        mass-metallicity relation of the ISM in their natal galaxies.
    \item More exotic mechanisms, such as minhalo or substructure collisions,
        may produce a small fraction of GCs ($1-2$ per cent), but this formation
        channel occurs rarely compared to more ``mundane'' mechanisms in the
        E-MOSAICS simulations.
\end{itemize}

These results present a parsimonious picture of GC formation. At high redshift,
the ISM of young galaxies frequently reached gas pressures high enough to form
the majority of GCs seen today.  These pressures were produced in turbulent,
gas-rich discs fed through cold accretion, and the clusters that survive today
are those that were ejected from their natal disc through mergers and interactions
during hierarchical galaxy assembly.  The birth environment of GCs is
the simplest one possible, namely the normal star-forming galaxies that were typical
during the epoch of GC formation.  As a result, the GCs in the E-MOSAICS
simulations are the relics of regular star formation in normal high-redshift galaxies. 

\section*{Acknowledgements}
The analysis was performed using pynbody (\texttt{http://pynbody.github.io/},
\citealt{pynbody}).   We also thank NSERC for funding supporting this research.
BWK and JMDK gratefully acknowledge funding from the European Research Council
(ERC) under the European Union's Horizon 2020 research and innovation programme
via the ERC Starting Grant MUSTANG (grant agreement number 714907).  JP and BP
acknowledge financial support from the European Research Council
(ERC-CoG-646928, Multi-Pop).  NB acknowledges financial support from the Royal
Society (University Research Fellowship). RAC is a Royal Society University
Research Fellow.  BWK acknowledges funding in the form of a Postdoctoral
Research Fellowship from the Alexander von Humboldt Stiftung. JMDK acknowledges
funding from the German Research Foundation (DFG) in the form of an Emmy Noether
Research Group (grant number KR4801/1-1). MRC is supported by a PhD Fellowship
from the International Max Planck Research School (IMPRS-HD).

\bibliographystyle{mnras}
\bibliography{references}
\appendix
\section{Comparison to the Li and Gnedin et al. Simulations}
\label{LiComparison}
The recent simulations presented in \citet{Li2017a}, \citet{Li2018}, and
\citet{Li2019a} (LG1719 henceforth) are similar in both methods and objectives
to the E-MOSAICS simulations we have used here, with a number of significant
differences.  Both E-MOSAICS and the LG1719 simulations are cosmological
zoom-ins of $L^*$ galaxies, simulated at resolutions higher than could be
obtained in comsological volumes with similar computation cost, but below the
extremely high resolution needed to resolve the individual stars within GCs.
Both simulations include sub-grid models for the birth, evolution, and
disruption of GCs, with key differences between the two models (described
below).  While the \citet{Li2017a} simulations are run at a comparable
resolution to the E-MOSAICS simulations, \citet{Li2018} and \citet{Li2019a}
increases their hydrodynamic resolution by a factor of 4, giving them somewhat
higher spatial resolution for the hydrodynamics than the E-MOSAICS simulations.
\footnote{It is somewhat non-trivial to directly compare hydrodynamic resolution
between Eulerian codes such as ART \citep{Kravtsov1997} (used by the LG1719
simulations) and Lagrangian ones such as ANARCHY-SPH \citep{Schaller2015}, used
for EAGLE and E-MOSAICS.  The LG1719 simulations use two refinement schemes, one
of which is quasi-Lagrangian, which gives an effective ``cell mass'' resolution
of $2.1\times10^5\Msun$, comparable to E-MOSAICS.  The highest spatial
hydrodynamic resolution of the \citet{Li2017a} simulations is 30 comoving pc,
compared to the minimum SPH smoothing length in E-MOSAICS of $35\pc$ after
$z=2.8$, and $133$ comoving pc prior to this.  In the re-simulations of
\citet{Li2018} and \citet{Li2019a}, the same root grid and quasi-Lagrangian
refinement scheme is used, giving an identical mass resolution, albeit with a
spatial resolution of 7.5 comoving pc.  Their use of a Jeans refinement criterion
is also similar to the use of an imposed equation of state used in E-MOSAICS to keep
the Jeans length above the resolution limit.} This additional resolution
naturally increases the cost of simulations, and as a result, the LG1719
simulations have only examined a single halo, with a single assembly history,
and only to a minimum redshift of $z=0.6$ (a look back time of $5.7\Gyr$).  As
we have shown in \citet{Pfeffer2018,Kruijssen2019a}, the different assembly
histories of
$L^*$ galaxies with similar masses can result in significantly different GC
populations from galaxy to galaxy.

Aside from the different hydrodynamic scheme, and the differences in the star
formation and stellar feedback models used in E-MOSAICS and LG1719, there are
some significant differences in the assumptions made in the star cluster models
between the two models.  The largest difference is in the formation model.
While E-MOSAICS uses a physically and observationally motivated cluster
formation efficiency and ICMF model to instantaneously build a population of
clusters, LG1719 builds clusters over a $15\Myr$ period of accretion, treating
newly-formed star particles as sinks which can accrete mass from their
birth environment for a brief period of time.  This allows the LG1719
simulations to examine
the origin of the CFE and ICMF, which is imposed by hand in E-MOSAICS.  The
stellar evolution model used in LG1719 is roughly comparable to the one used in
E-MOSAICS (\citet{Conroy2010} and \citet{Wiersma2009} respectively).  However,
the tidal disruption models used in E-MOSAICS and LG1719 are quite different.
Both models rely on a local calculation of the tidal tensor:
\begin{equation}
    T_{ij} = \frac{\partial\Phi}{\partial x_i\partial x_j}
    \label{tidaltensor}
\end{equation}
and in particular the eigenvalues of the tidal tensor $\lambda_i$.  E-MOSAICS
and LG1719 both use these eigenvalues to determine the rate of mass loss due to
two-body relaxation (equation 13 in \citealt{Pfeffer2018} and equation 4
in \citealt{Li2019a}).  There is little difference in the equations used to
calculate these rates, with E-MOSAICS using:
\begin{equation}
    \left(\frac{dM}{dt}\right)_{ev} =
    4.7\times10^{-2}\Msun\Myr^{-1}\left(\frac{M}{\Msun}\right)^{-0.38}
    \left(\frac{T}{T_\odot}\right)^{1/2}
    \label{mosaicsdmdt}
\end{equation}
and LG1719 using:
\begin{equation}
    \left(\frac{dM}{dt}\right)_{ev} =
    5.9\times10^{-2}\Msun\Myr^{-1}\left(\frac{M}{\Msun}\right)^{-1/3}
    \left(\frac{T}{T_\odot}\right)^{1/2}
    \label{lgdmdt}
\end{equation}
As was explored in \citet{Pfeffer2018}, the slight changes in normalisation and
the exponent of the mass term produce very little difference in the overall
relaxation rate, and roughly correspond to the change brought by assuming a
different cluster density profile.   However, the equations used for estimating
the strength of the tidal field $T$ is quite different.  In E-MOSAICS, the tidal
field strength is estimated as:
\begin{equation}
    T = \max(\lambda_i) - \frac{1}{3}\sum_i \lambda_i
    \label{mosaicsT}
\end{equation}
While the LG1719 simulations use instead:
\begin{equation}
    T = \max(|\lambda_i|) 
    \label{LGT}
\end{equation}
Notably, this omits the term due to the Coriolis force $\Omega^2 =
\frac{1}{3}\sum_i \lambda_i$.  \citet{Li2019a} argues that this term is
unimportant for their clusters, based on estimates on the rotational velocity
and size of their high-redshift discs.  However, as figures C1 and C2 of
\citet{Pfeffer2018} shows, the circular frequency term $\Omega^2$ contributes
significantly to the tidal field in the inner $5\kpc$ of the $z=0$ disc, and
omitting it can lead to erroneously strong compressive tidal fields.
\citet{Li2019a} explicitly allows compressive tides to drive cluster
evaporation by taking the magnitude of $\lambda_i$, while E-MOSAICS does not.
This likely means that the mass loss rate due to tidal evaporation is higher in
the LG1719 simulations compared to the E-MOSAICS simulations.  

\begin{figure}
    \includegraphics[width=0.5\textwidth]{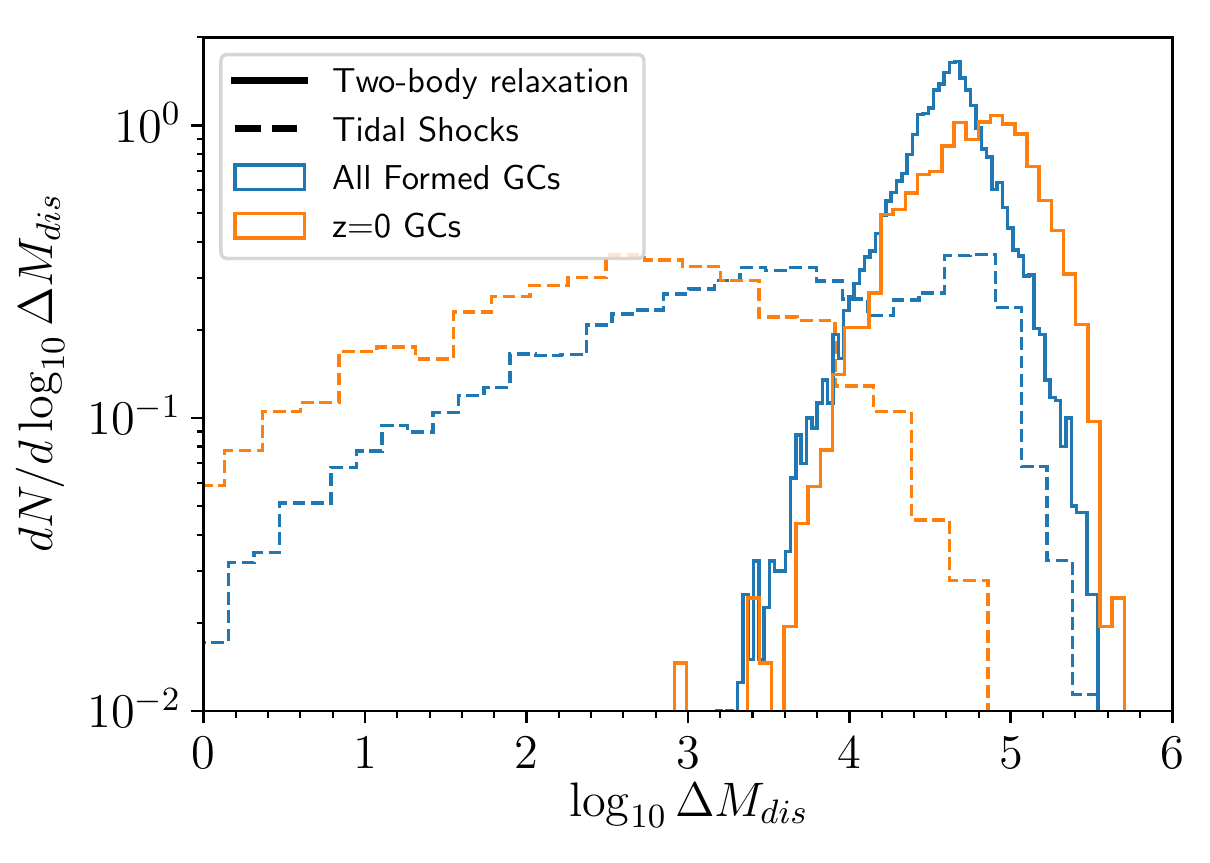}
    \caption{Integrated mass loss by disruption for all globular clusters
    observed at $z=0$ and all proto-GCs which form (including those which are
    disrupted by $z=0$).  As is clear, the $z=0$ population of GCs have
    experienced much more mass loss through two-body relaxation than through
    tidal shocks.  However, the population of {\it all} proto-GCs experience a
    nearly equal contribution from two-body relaxation and by tidal shocks.
    Omitting the effects of tidal shocks will result not only in more GCs
    surviving until $z=0$, but a different population of GCs compared to that
    which would be seen when including the effect of tidal shocks.}
    \label{shock_disruption}
\end{figure}

There is, however, another major difference in the calculated mass loss rates
that likely outweighs this effect: LG1719 omits a model for tidal shocks. Past
studies \citep{Spitzer1958,Ostriker1972,Kundic1995,Gnedin1997} have shown that
tidal shocks can transfer significant kinetic energy to stellar clusters, and
their effect was shown explicitly in \citet{Kruijssen2011} to contribute 80-85
per cent of all cluster disruption.  The E-MOSAICS simulations use the same
equations to determine the shock disruption rate as were used in
\citet{Kruijssen2011}.  It was shown in \citet{Kruijssen2012b} that tidal shocks
during major mergers can destroy more clusters than are formed during the
increased star formation induced by the merger.  As
Figure~\ref{shock_disruption} shows, the contribution of tidal shocks to the
disruption of proto-GCs  is significant, with two-body relaxation contributing
an average mass loss of $6.7\times10^4\Msun$ and tidal shocks contributing an
average mass loss of $4.9\times10^4\Msun$.  Not only is the relative
contribution from tidal shocks and two-body relaxation roughly equivalent in
proto-GCs, the difference between the $z=0$ GC population and the population of
proto-GCs shows that essentially no GCs surviving to $z=0$ have experienced mass
loss from tidal shocks that exceeds $10^5\Msun$, while the mass loss experienced
from both populations by two-body relaxation is relatively similar.  This
suggests tidal shocks play an important role in establishing which proto-GCs
survive to $z=0$.  Those clusters we see surviving to $z=0$ are the ones which
have been relatively unaffected by tidal shocks.  

The effects of dynamical friction are also omitted from the LG1719 simulations,
which is applied as a post-processing treatment in E-MOSAICS following
\citet{Lacey1993}.  Interestingly, despite the differences in both the formation
model and the treatment of tidal disruption, both E-MOSAICS and LG1719 produce a
similar final CMF, with an overabundance of low mass clusters.  This suggests
that the similar treatments of tidal disruption may need to be improved in the
future to increase the disruption rate of low mass clusters.  Despite the
similarity in the CMFs in E-MOSAICS and LG1719, the metallicity distributions
show opposite issues: too many metal-poor clusters in LG1719 and too many
metal-rich clusters in E-MOSAICS.  The different numerical approaches are the
likely explanation of this.  In E-MOSAICS, the ISM is under-resolved due to the
fixed Jeans equation of state, leading to under-disruption in the metal-rich
disc at lower redshifts \citep{Kruijssen2019a}.  Meanwhile, in LG1719, tidal
shocks are not included in the disruption rates, leading to under-disruption of
globular clusters forming in high-redshift, low metallicity galaxies which
experience frequent, violent mergers and inflow driven turbulence.  With future
improvements to the treatment of cold gas in E-MOSAICS and tidal disruption in
LG1719, it is likely that both of these issues will be resolved and the results
from the two simulation sets will further converge.

\end{document}